\documentclass[english]{article}
\usepackage[T1]{fontenc}
\usepackage[latin9]{inputenc}
\usepackage{xcolor}
\usepackage{pdfcolmk}
\usepackage{url}
\usepackage{amsmath}
\usepackage{amsthm}
\usepackage{amssymb}
\usepackage{graphicx}
\usepackage[authoryear]{natbib}
\PassOptionsToPackage{normalem}{ulem}
\usepackage{ulem}

\makeatletter

\providecolor{lyxadded}{rgb}{0,0,1}
\providecolor{lyxdeleted}{rgb}{1,0,0}

\DeclareRobustCommand{\lyxsout}[1]{\ifx\\#1\else\sout{#1}\fi}

\usepackage[preprint]{neurips_2019}

\usepackage{hyperref}
\usepackage{url}

\usepackage{ccaption}

\@ifundefined{showcaptionsetup}{}{%
 \PassOptionsToPackage{caption=false}{subfig}}
\usepackage{subfig}
\makeatother

\usepackage{babel}
\begin{document}
\author{
Alexander Immer \\
EPFL \\
\texttt{alexander.immer@epfl.ch} \\
\And
Guillaume Dehaene \\
EPFL \\
\texttt{guillaume.dehaene@gmail.com} \\
}
\title{Variational Inference with Numerical Derivatives: variance reduction
through coupling}
\maketitle
\begin{abstract}
The Black Box Variational Inference (\citet{ranganath2014black})
algorithm provides a universal method for Variational Inference, but
taking advantage of special properties of the approximation family
or of the target can improve the convergence speed significantly.
For example, if the approximation family is a transformation family,
such as a Gaussian, then switching to the reparameterization gradient
(\citet{kingma2013auto}) often yields a major reduction in gradient
variance. Ultimately, reducing the variance can reduce the computational
cost and yield better approximations.

We present a new method to extend the reparameterization trick to
more general exponential families including the Wishart, Gamma, and
Student distributions. Variational Inference with Numerical Derivatives
(VIND) approximates the gradient with numerical derivatives and reduces
its variance using a tight coupling of the approximation family. The
resulting algorithm is simple to implement and can profit from widely
known couplings. Our experiments confirm that VIND effectively decreases
the gradient variance and therefore improves the posterior approximation
in relevant cases. It thus provides an efficient yet simple Variational
Inference method for computing non-Gaussian approximations.
\end{abstract}

\section{Introduction}

Variational methods offer a promising path towards making Bayesian
inference feasible for fitting large models to large datasets (\citet{blei2017variational}).
Indeed, a generic strategy exists to approximate a posterior distribution
inside any parametric family. \citet{ranganath2014black} give the
BBVI (Black Box Variational Inference) formula for a stochastic gradient
of the ELBO which can then be plugged into any stochastic optimization
algorithm. However, replacing this generic stochastic gradient by
taking advantage of the structure of the approximating can yield a
considerable reduction of the variance thus speeding up the convergence
considerably. For example, when the approximating family is a transformation
family (e.g. Gaussian approximations), the ``reparameterization''
stochastic gradient (\citet{kingma2013auto}) can be used instead.

In this article, we propose an extension of the reparameterization
gradient to more general families. Our algorithm, VIND (Variational
Inference through Numerical Derivatives) approximates the gradient
using finite differences. By ``coupling'' or correlating the two
approximations at which we compute the ELBO, we achieve a large reduction
in variance compared to the BBVI gradient thus considerably improving
the convergence. We demonstrate how to derive such couplings for Gamma,
Beta, Dirichlet, Wishart, univariate and multivariate Student and
Poisson approximations of the posterior distribution (Appendix Section
\ref{sec:Couplings}). In our experiments\footnote{Code available at \url{https://github.com/AlexImmer/VIND}.},
the VIND gradient improves upon the BBVI gradient. Indeed, we found
that the BBVI gradient is unable to optimize the degree-of-freedom
parameter of a Wishart approximation while VIND efficiently finds
the optimal value.

\section{Background}

Throughout this article, we will consider the problem of computing
a variational approximation of a target probability density: $f\left(\boldsymbol{\theta}\right)$.
We seek to find the distribution which, inside a parametric family
$q\left(\boldsymbol{\theta};\boldsymbol{\lambda}\right)$, maximizes
the Evidence Lower BOund (ELBO). This yields both the best ELBO-based
approximation of the normalization constant of $f\left(\boldsymbol{\theta}\right)$
and the best approximation of the normalized density $f\left(\boldsymbol{\theta}\right)$
within the $q\left(\boldsymbol{\theta};\boldsymbol{\lambda}\right)$
family, as measured by the KL divergence (\citet{blei2017variational}).
The ELBO has two alternative expressions:
\begin{align}
ELBO & =\mathbb{E}_{\boldsymbol{\theta}\sim q\left(\boldsymbol{\theta};\boldsymbol{\lambda}\right)}\left[\log\frac{p\left(\boldsymbol{\theta}\right)}{q\left(\boldsymbol{\theta};\boldsymbol{\lambda}\right)}\right]\label{eq: ELBO definition}\\
 & =\mathbb{E}_{\boldsymbol{\theta}\sim q\left(\boldsymbol{\theta};\boldsymbol{\lambda}\right)}\left[\log p\left(\boldsymbol{\theta}\right)\right]+H\left(\boldsymbol{\lambda}\right)\label{eq: ELBO with entropy}
\end{align}
where $H\left(\boldsymbol{\lambda}\right)$ is the entropy of a random
variable with density $q\left(\boldsymbol{\theta};\boldsymbol{\lambda}\right)$.

The ELBO can always be optimized through Stochastic Gradient Descent.
Indeed, the gradient of the ELBO with respect to $\boldsymbol{\lambda}$
is (\citet{ranganath2014black}):
\begin{equation}
\nabla_{\boldsymbol{\lambda}}ELBO=\mathbb{E}_{\boldsymbol{\theta}\sim q\left(\boldsymbol{\theta};\boldsymbol{\lambda}\right)}\left[\log\frac{p\left(\boldsymbol{\theta}\right)}{q\left(\boldsymbol{\theta};\boldsymbol{\lambda}\right)}\nabla_{\boldsymbol{\lambda}}\left\{ \log q\left(\boldsymbol{\theta};\boldsymbol{\lambda}\right)\right\} \right]\label{eq: grad ELBO general form}
\end{equation}
This expected value can be approximated by sampling from $\boldsymbol{\theta}_{\boldsymbol{\lambda}}\sim q\left(\boldsymbol{\theta};\boldsymbol{\lambda}\right)$.
The resulting stochastic optimization algorithm is referred to as
Black Box Variational Inference (BBVI) to highlight the absence of
any condition on the approximating family.

However, we can often take advantage of the structure of the approximation
family. One particularly nice case is when we consider a transformation
family, i.e. when the variable $\boldsymbol{\theta}_{\boldsymbol{\lambda}}$
with distribution $q\left(\boldsymbol{\theta};\boldsymbol{\lambda}\right)$
can be rewritten as a deterministic transformation of a base random
variable $\boldsymbol{Z}$:
\begin{equation}
\boldsymbol{\theta}_{\boldsymbol{\lambda}}=F\left(\boldsymbol{Z},\boldsymbol{\lambda}\right)
\end{equation}

For example, all Gaussians can be rewritten as an affine transformation
of a standard Gaussian. This structure greatly simplifies the evaluation
of the gradient of the ELBO. Rewriting the ELBO as an expected value
under $\boldsymbol{Z}$ and denoting the Jacobian matrix with $J_{\boldsymbol{\lambda}}$,
the gradient against $\boldsymbol{\lambda}$ becomes:
\begin{equation}
\nabla_{\boldsymbol{\lambda}}ELBO=\mathbb{E}\left[J_{\boldsymbol{\lambda}}\left\{ F\left(\boldsymbol{Z},\boldsymbol{\lambda}\right)\right\} \nabla_{\boldsymbol{\theta}}\left\{ \log\frac{p\left(F\left(\boldsymbol{Z},\boldsymbol{\lambda}\right)\right)}{q\left(F\left(\boldsymbol{Z},\boldsymbol{\lambda}\right);\boldsymbol{\lambda}\right)}\right\} \right]\label{eq: grad ELBO reparameterization form}
\end{equation}
which can once again be evaluated by sampling from $\boldsymbol{Z}$.
Critically, shifting from the general form (eq.\eqref{eq: grad ELBO general form})
to the reparameterization form of the gradient (eq.\eqref{eq: grad ELBO reparameterization form})
results in a major decrease of variance (\citet{kingma2013auto,ruiz2016generalized}).
Intuitively, this makes sense since we have replaced an optimization
over the space of probability distributions (eq.\eqref{eq: grad ELBO general form})
by an optimization over deterministic transformations of $\boldsymbol{Z}$.
(eq.\eqref{eq: grad ELBO reparameterization form}).

However, many approximation families are not amenable to a full reparameterization.
For example, Gamma distributions:
\begin{equation}
q\left(\theta;\alpha,\beta\right)\propto\theta^{\alpha}\exp\left(-\beta\theta\right)
\end{equation}
are only partially reparameterizable: the $\beta$ parameter is an
inverse-scale and can thus be absorbed into a linear reparameterization
but it is impossible to do so for the $\alpha$ parameter. For families
with such parameters, we propose to replace BBVI by a finite difference
scheme, while doing a standard reparameterization gradient on all
other parameters. By considering a sampling approximation of the ELBO
at $\alpha-\epsilon$ and $\alpha+\epsilon$, we obtain an alternative
to the BBVI ELBO gradient (eq.\eqref{eq: grad ELBO general form}):
\begin{align}
\nabla_{\alpha}ELBO\left(\alpha,\beta\right) & \approx\frac{1}{2\epsilon}\Bigg(\mathbb{E}_{\theta\sim q\left(\theta;\alpha+\epsilon,\beta\right)}\left[\log\left\{ \frac{p\left(\theta\right)}{q\left(\theta;\alpha,\beta\right)}\right\} \right]\nonumber \\
 & \ \ \ \ -\mathbb{E}_{\theta\sim q\left(\theta;\alpha-\epsilon,\beta\right)}\left[\log\left\{ \frac{p\left(\theta\right)}{q\left(\theta;\alpha,\beta\right)}\right\} \right]\Bigg)
\end{align}
 However, independent sampling from $q\left(\theta;\alpha-\epsilon,\beta\right)$
and $q\left(\theta;\alpha+\epsilon,\beta\right)$ would yield a larger
variance than necessary. Instead, we propose to sample from a coupling
of these two distributions: a heavily correlated joint distribution
$q\left(\theta_{\alpha+\epsilon},\theta_{\alpha-\epsilon}\right)$
such that its marginals are $q\left(\theta;\alpha+\epsilon,\beta\right)$
and $q\left(\theta;\alpha-\epsilon,\beta\right)$ (\citet{thorisson1995coupling,villani2008optimal};
see \citet{propp1996exact} for an application to MCMC sampling).
Comparison between $\alpha-\epsilon$ and $\alpha+\epsilon$ is then
easy: it suffices to compare the empirical mean under the $\theta_{\alpha+\epsilon}$
samples to the mean under the $\theta_{\alpha-\epsilon}$ samples.

\citet{ruiz2016generalized} propose instead to tackle such parameters
by performing an approximate reparameterization. More precisely, they
propose to standardize $\boldsymbol{\theta}_{\lambda}$ (or a transformation
of it) so that it has mean $0$ and covariance the identity matrix.
For example, for the Gamma approximation, they propose to standardize
$\log\left(\theta\right)$:
\begin{equation}
Z|\alpha=\frac{\log\left(\theta\right)-\mathbb{E}\left(\log\left(\theta\right);\alpha,\beta\right)}{\sqrt{\text{Var}\left(\log\left(\theta\right);\alpha,\beta\right)}}
\end{equation}
The resulting standardized variable has a distribution that does not
depend on $\beta$ and has a weaker relationship to $\alpha$ than
$\log\left(\theta\right)$: the conditional mean of $Z$ is constant
while the conditional mean of $\log\left(\theta\right)$ is roughly
equal to $\log\left(\alpha\right)$. The gradient of the ELBO then
decomposes into a reparameterization term and a BBVI remainder term
and has a smaller variance than the standard BBVI gradient (\citet{ruiz2016generalized}).
This Generalized Reparameterization gradient (GREP) is directly parallel
to VIND in that it tries to extend the good properties of the reparameterization
gradient to families which do not possess a transformation family
structure. However, one sharp limit of GREP is that it is highly non-trivial
to derive appropriate standardizations of a given approximation family.

Another possibility to minimize the variance consists of using control
variates (see \citet{geffner2018using} and the references therein).
This approach minimizes the variance by identifying functions of $\boldsymbol{\theta}$
which have a known expected value and which can closely approximate
the log-ratio $\log\frac{p\left(\boldsymbol{\theta}\right)}{q\left(\boldsymbol{\theta};\boldsymbol{\lambda}\right)}$.
The ELBO can then be decomposed into a deterministic term and smaller
stochastic remainder thus minimizing the variance. While control variates
approaches are necessary to get the optimal variant of a stochastic
optimization method, it is orthogonal to ideas such as VIND which
aim to radically modify the form of the gradient. We will investigate
in further work how to properly combine VIND with control-variates
methods in order to achieve maximal efficiency.

\section{Variational Inference with Numerical Derivatives}

Computing gradients with finite differences is simple and intuitive.
We just vary the parameter $\lambda$ by $\pm\epsilon$ while computing
the loss to get an approximation of the gradient. This yields an approximation
of the gradient that is accurate up to $\mathcal{O}\left(\epsilon^{2}\right)$
errors. For example, if the parameter $\lambda$ is one dimensional,
we can approximate the gradient as:
\begin{align}
\widetilde{\nabla}_{\lambda}ELBO & =\frac{ELBO(\lambda+\epsilon)-ELBO(\lambda-\epsilon)}{2\epsilon}\\
 & =\frac{1}{2\epsilon}\mathbb{E}\left[\log\frac{p\left(\theta_{\lambda+\epsilon}\right)}{q\left(\theta_{\lambda+\epsilon};\lambda+\epsilon\right)}-\log\frac{p\left(\theta_{\lambda-\epsilon}\right)}{q\left(\theta_{\lambda-\epsilon};\lambda-\epsilon\right)}\right]\label{eq: Finite differences approximation of the gradient}
\end{align}
where $\theta_{\lambda\pm\epsilon}$ are random variables with distribution
$q\left(\theta;\lambda\pm\epsilon\right)$. The VIND update is a slight
refinement of eq.\eqref{eq: Finite differences approximation of the gradient}
which computes the density of the approximation $q$ at parameter
value $\lambda$ instead of $\lambda\pm\epsilon$:
\begin{equation}
\nabla_{\lambda,VIND}ELBO=\frac{1}{2\epsilon}\mathbb{E}\left[\log\frac{p\left(\theta_{\lambda+\epsilon}\right)}{q\left(\theta_{\lambda+\epsilon};\lambda\right)}-\log\frac{p\left(\theta_{\lambda-\epsilon}\right)}{q\left(\theta_{\lambda-\epsilon};\lambda\right)}\right]\label{eq: VIND gradient}
\end{equation}
This modification ignores a term with gradient equal to $0$. Ignoring
this term thus reduces the variance while still yielding an approximation
of the gradient (Appendix Section \ref{sec:Derivation-of-the VIND}).
Please notice that eq.\eqref{eq: VIND gradient} requires that both
values $\lambda\pm\epsilon$ correspond to valid parameterizations
so that we can sample from both $\theta_{\lambda+\epsilon}$ and $\theta_{\lambda-\epsilon}$.
This might require tuning the value of $\epsilon$ to $\lambda$ or
to perform one-sided finite-differences approximations instead when
$\lambda$ is close to the edges of the parameter space.

As in BBVI and the reparameterized form of the ELBO, the gradient
approximation (eq.\eqref{eq: VIND gradient}) can be evaluated through
sampling from $q\left(\boldsymbol{\theta};\lambda\pm\epsilon\right)$.
This is non-trivial as sampling independently from these two distributions
generally yields a high-variance gradient estimator (Appendix Fig.\ref{fig:Estimated-variance-of VIND, VIND_uncoupled, BBVI}).
The key trick consists instead of sampling from a \emph{coupling}
(\citet{thorisson1995coupling,villani2008optimal}):\emph{ }a joint
distribution $q\left(\theta_{\lambda+\epsilon},\theta_{\lambda-\epsilon}\right)$
such that $\theta_{\lambda+\epsilon}$ and $\theta_{\lambda-\epsilon}$
are marginally respectively distributed from $q\left(\theta;\lambda\pm\epsilon\right)$.
There are infinitely many such couplings with the independent coupling
being the simplest one. For a finite-difference approximation of the
gradient, it is more efficient for $\theta_{\lambda+\epsilon},\theta_{\lambda-\epsilon}$
to have high-positive correlation. Indeed, consider the term $\mathbb{E}_{\theta}\left[\log\left\{ p\left(\theta\right)\right\} \right]$
in the ELBO. The variance of the corresponding approximation of the
gradient is:

\begin{align}
\text{Var}\left[\frac{\log p(\theta_{\lambda+\epsilon})-\log p(\theta_{\lambda-\epsilon})}{2\epsilon}\right] & =\text{Var}\left[\frac{\log p(\theta_{\lambda+\epsilon})}{2\epsilon}\right]+\text{Var}\left[\frac{\log p(\theta_{\lambda-\epsilon})}{2\epsilon}\right]\nonumber \\
 & -2\text{Cov}\left[\frac{\log p(\theta_{\lambda+\epsilon})}{2\epsilon},\frac{\log p(\theta_{\lambda-\epsilon})}{2\epsilon}\right]
\end{align}
which is minimized when $\log p\left(\theta_{\lambda+\epsilon}\right)$
and $\log p\left(\theta_{\lambda-\epsilon}\right)$ are highly positively
correlated.

The variance of the gradient determines the convergence speed and
the possible obtainable minimum loss (\citet{bottou-98x,duchi2011adaptive,kingma2014adam}).
In VIND, the variance depends on our choice of $\text{\ensuremath{\epsilon}}$
and the particular coupling that we use. Larger values of $\epsilon$
typically lead to smaller variance but they also lead to the numerical
approximation of the derivative of the ELBO becoming rougher. It is
not \emph{a priori} clear which choice is optimal. Our experiments
indicate that the range of values of acceptable values of $\epsilon$
is large (Fig. \ref{fig: 1 MSE}).

\subsection{Coupling the Gamma distribution}

Let us now show how we can achieve a coupling with high covariance
for the Gamma distribution (see Appendix Section \ref{sec:Couplings}
for coupling of other families). The family of Gamma distributions
has a key property: the sum of two Gamma variables with the same $\beta$
is still a Gamma variable. Mathematically, for any $\alpha_{1},\alpha_{2},\beta>0$,
we have: \begin{subequations}
\begin{align}
X & \sim\Gamma(\alpha_{1},\beta)\\
Y & \sim\Gamma(\alpha_{2},\beta)\\
X+Y & \sim\Gamma(\alpha_{1}+\alpha_{2},\beta)
\end{align}
\end{subequations} Furthermore, the $\beta$ parameter is amenable
to reparameterization since it is a inverse-scale parameter. It is
thus straightforward to express to construct a coupling of $q\left(\theta;\alpha\pm\epsilon,\beta\right)$,
reparameterized in $\beta$, using three auxiliary Gamma variables
as:\begin{subequations}
\begin{align}
\gamma_{\alpha-\epsilon} & \sim\Gamma\left(\alpha-\epsilon,1\right)\\
\gamma_{\epsilon,1} & \sim\Gamma\left(\epsilon,1\right)\\
\gamma_{\epsilon,2} & \sim\Gamma\left(\epsilon,1\right)\\
\theta_{\alpha-\epsilon,\beta} & =\frac{1}{\beta}\gamma_{\alpha-\epsilon}\\
\theta_{\alpha+\epsilon,\beta} & =\frac{1}{\beta}\left(\gamma_{\alpha-\epsilon}+\gamma_{\epsilon,1}+\gamma_{\epsilon,2}\right)\\
\theta_{\alpha,\beta} & =\frac{1}{\beta}\left(\gamma_{\alpha-\epsilon}+\gamma_{\epsilon,1}\right)
\end{align}
\end{subequations}

Given $n$ independent samples from this coupling $\theta^{\left(1\right)}\dots\theta^{\left(n\right)}$,
we can evaluate the gradient of the ELBO against $\beta$ with the
reparameterization formula (eq.\eqref{eq: grad ELBO reparameterization form}):
\begin{equation}
\nabla_{\beta}ELBO\approx\frac{1}{n}\left\{ \sum_{j=1}^{n}\left(-\frac{\gamma_{\alpha-\epsilon}+\gamma_{\epsilon,1}}{\beta^{2}}\right)\frac{\partial}{\partial\theta}\left[\log\frac{p\left(\theta_{\alpha,\beta}^{\left(j\right)}\right)}{q\left(\theta_{\alpha,\beta}^{\left(j\right)};\alpha,\beta\right)}\right]\right\} 
\end{equation}
while the gradient against $\alpha$ can be approximated through the
VIND formula (eq.\eqref{eq: VIND gradient}):
\begin{equation}
\nabla_{\alpha,VIND}ELBO\approx\frac{1}{2\epsilon}\frac{1}{n}\sum_{j=1}^{n}\left(\log\frac{p\left(\theta_{\alpha+\epsilon,\beta}^{\left(j\right)}\right)}{q\left(\theta_{\alpha+\epsilon,\beta}^{\left(j\right)};\alpha,\beta\right)}-\log\frac{p\left(\theta_{\alpha-\epsilon,\beta}^{\left(j\right)}\right)}{q\left(\theta_{\alpha-\epsilon,\beta}^{\left(j\right)};\alpha,\beta\right)}\right)
\end{equation}

\subsection{General formulation}

The general procedure of the VIND algorithm is as follows. In order
to compute a variational approximation of $f\left(\boldsymbol{\theta}\right)$
inside a given parametric family $q\left(\boldsymbol{\theta};\boldsymbol{\lambda}\right)$,
we need to first identify all the parameters that are not amenable
to normal reparameterization and for which we desire to compute a
VIND gradient. For every such parameter $\lambda_{i}$, we need to
construct a joint distribution of $\boldsymbol{\theta}_{\dots\lambda_{i}+\epsilon\dots}$
and $\boldsymbol{\theta}_{\dots\lambda_{i}-\epsilon\dots}$ where
the $i^{th}$ coordinate of $\boldsymbol{\lambda}$ has been perturbed
by epsilon. Finally, the VIND approximation of the gradient to $\lambda_{i}$
is constructed and then approximated using $n$ samples from the coupling
$\boldsymbol{\theta}^{\left(1\right)}\dots\boldsymbol{\theta}^{\left(n\right)}$:
\begin{align}
\nabla_{\lambda_{i},VIND}ELBO & =\frac{1}{2\epsilon}\mathbb{E}\left[\log\frac{p\left(\boldsymbol{\theta}_{\dots\lambda_{i}+\epsilon\dots}^{\left(j\right)}\right)}{q\left(\boldsymbol{\theta}_{\dots\lambda_{i}+\epsilon\dots}^{\left(j\right)};\boldsymbol{\lambda}\right)}-\log\frac{p\left(\boldsymbol{\theta}_{\dots\lambda_{i}-\epsilon\dots}^{\left(j\right)}\right)}{q\left(\boldsymbol{\theta}_{\dots\lambda_{i}-\epsilon\dots}^{\left(j\right)};\boldsymbol{\lambda}\right)}\right]\label{eq: the VIND update}\\
\nabla_{\lambda_{i},VIND}ELBO & \approx\frac{1}{2\epsilon}\frac{1}{n}\sum_{j=1}^{n}\left[\log\frac{p\left(\boldsymbol{\theta}_{\dots\lambda_{i}+\epsilon\dots}^{\left(j\right)}\right)}{q\left(\boldsymbol{\theta}_{\dots\lambda_{i}+\epsilon\dots}^{\left(j\right)};\boldsymbol{\lambda}\right)}-\log\frac{p\left(\boldsymbol{\theta}_{\dots\lambda_{i}-\epsilon\dots}^{\left(j\right)}\right)}{q\left(\boldsymbol{\theta}_{\dots\lambda_{i}-\epsilon\dots}^{\left(j\right)};\boldsymbol{\lambda}\right)}\right]
\end{align}

Thus, in order to extend this approach to other approximation families,
we need to devise an appropriate coupling. In the Appendix, we give
couplings for the Gamma, Beta, Dirichlet, Wishart, univariate and
multivariate Student and Poisson distributions (Appendix Section \ref{sec:Couplings}).

The VIND algorithm generalizes the reparameterization gradient. Indeed,
transformation families have a straightforward coupling which is such
that the VIND gradient coincides to the reparameterization gradient
in the limit $\epsilon\rightarrow0$ (Appendix Section \ref{sec:Linking-VIND-to}).
These two methods should thus be similar. While we have not been able
to provide theoretical reasons for the VIND gradient to have a smaller
variance than the BBVI gradient, we believe that the following heuristic
explanation might provide some insight. The reparameterization gradient
is able to improve upon BBVI by providing a simpler comparison between
the ELBO at various values of $\boldsymbol{\lambda}$. Instead of
comparing how the expected value under $q\left(\boldsymbol{\theta};\boldsymbol{\lambda}\right)$
is modified, we seek instead the best deterministic transformation
of a base variable $\boldsymbol{Z}$ through the mapping $F\left(\boldsymbol{Z};\boldsymbol{\lambda}\right)$.
This modifies the space on which we are optimizing from the complicated
space of probability distributions to the simpler space of deterministic
deformations of a base probability distribution. The VIND idea expands
upon this idea of finding the best deformation except we further consider
stochastic deformations in order to deal with families which do not
have the transformation family structure.

\section{Experiments}

We apply VIND to both synthetic and real-world problems and compare
it to the standard BBVI approach (\citet{ranganath2014black}) and
the generalized reparameterization gradient of \citet{ruiz2016generalized}
(GREP). We use synthetic data and a conjugate model to showcase the
variance reduction compared to BBVI and highlight the influence of
different values of the hyperparameter $\epsilon$. Further, we compare
VIND to BBVI on a covariance estimation problem for financial data
where the approach of \citet{ruiz2016generalized} cannot be applied.
Throughout this section, we use Adam as the gradient descent routine
with standard parameters (\citet{kingma2014adam}) but adjusted learning
rates per parameter. In Appendix Section \ref{subsec:Further-Experiments},
we provide additional details and an additional application to linear
regression.

\subsection{Gradient MSE in a Gamma-Normal model\label{sec: synthetic data}}

To investigate the mean-squared error of the VIND gradient estimator,
we use a synthetic set of data $\mathcal{D}=\{x_{i}\}_{i=1}^{n}$
sampled from a univariate Gaussian distribution of known mean $0$,
i.e. $x_{i}\sim\mathcal{N}(0,\tau^{-1})$. We want to find the posterior
over $\tau$. The conditional distribution is a Gaussian with fixed
mean and the prior is a Gamma distribution. Due to conjugacy of this
model, we have access to the actual gradient and can estimate the
variance and bias of the VIND and BBVI gradients. The conditional
model and prior are defined as follows:

\begin{equation}
f(\mathcal{D}|\tau)=\prod_{i=1}^{n}\mathcal{N}(x_{i}|\mu,\tau^{-1}),\qquad p(\tau)=\Gamma(\tau;\alpha_{0},\beta_{0})
\end{equation}

We approximate the true posterior $f(\tau|\mathcal{D})$ with a Gamma
distribution $q(\tau;\alpha)$. $\alpha$ is the variational parameter
to fit while the rate $\beta$ is set to the posterior parameter according
to the conjugate model. This is done to focus on the $\alpha$ parameter
which cannot be optimized through a reparameterization gradient. We
optimize the ELBO using the true gradient. At each step, the gradient
mean-squared error (MSE) is estimated for VIND with different values
of $\epsilon$ and BBVI. Both methods use two samples per gradient
estimate.

\begin{figure}
\centering{}\includegraphics[width=0.6\textwidth]{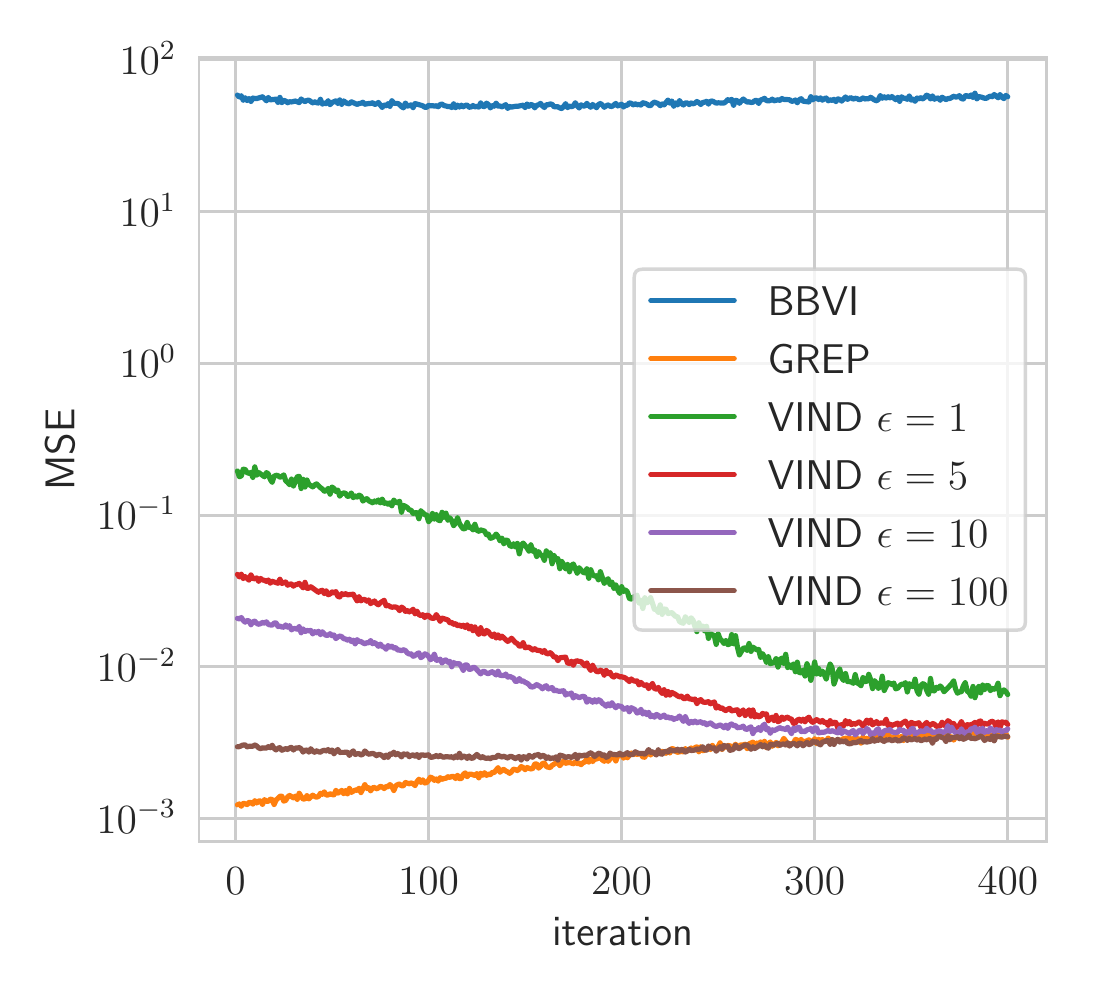}\caption{\textbf{Gradient MSE comparison in the Gamma-normal model.pdf}\protect \\
Log-scaled plot of estimated MSE of different VIND-$\epsilon$ gradients,
GREP gradient (\citet{ruiz2016generalized}) and BBVI gradient along
the ELBO ascent path using the true gradient. All gradient estimators
use two samples. The MSE of the VIND estimators is consistently lower
than BBVI except for the smallest value $(\epsilon=0.1)$, where it
is initially the same. The GREP gradient is optimal in this experiment
but the VIND gradient is competitive in later iterations. The variance
dominates the MSE while the bias is small for all estimators, even
for $\epsilon=10$ (see Appendix Section \ref{subsec:Synthetic-Gamma-Normal}).\label{fig: 1 MSE}}
\end{figure}

Fig.\ref{fig: 1 MSE} depicts the MSE of different gradient estimators.
The gradient obtained using VIND yields a lower MSE for a wide range
of values for $\epsilon$ compared to the BBVI gradient. This experiment
shows that the VIND update provides a massive reduction in variance
compared to BBVI, comparable to that achieved by \citet{ruiz2016generalized}
while the bias is negligible compared to the variance.

\subsection{Covariance Estimation for Financial Data}

We further considered a data set of weekly log returns of the Dow
Jones in the period from 2011 to 2018. Data is obtained for $n=400$
weeks and $d=29$ of the 30 companies constituting the Dow Jones as
2018 (for a list of stock symbols see Appendix Section \ref{subsec:Wishart-Student-Financial})\footnote{The historical data are obtained from \url{https://finance.yahoo.com}.}.
We denote the data set by $\mathcal{D}=\{\boldsymbol{x}_{i}\}_{i=1}^{n}$
with $\boldsymbol{x}_{i}\in\mathbb{R}^{d}$. The data pass a stationarity
and autocorrelation test, i.e. it is reasonable to assume fixed mean
and variance. We conduct a wavelet spectrum test to account for second
order stationarity (\citet{nason2013test}) and the Durbin-Watson
test to exclude autocorrelation (\citet{durbin1950testing,durbin1951testing}).

To apply portfolio optimization models that trade off risk and return,
both mean and covariance of the options need to be estimated (\citet{ryan2006portfolio}).
Bayesian methods have the advantage over frequentist methods that
we have access to the parameter uncertainty. However, the only conjugate
model available assumes the data to be distributed according to a
Gaussian. For a more elaborate model, one therefore usually resorts
to approximate inference.

We treat this as a classical Bayesian multivariate location-scale
estimation problem. For the conditional, we choose a t-Student distribution
(i.e. the marginal distribution of a multivariate Gaussian variable
$\mathcal{N}\left(0,\Sigma\right)$ divided by a $\chi_{\nu}^{2}$
variable, where $\nu$ is the degree of freedom parameter of the Student)
modeling the possibility of major deviations from the mean that are
almost impossible under a Gaussian model. The additional degree of
freedom parameter allows fitting the shape more closely. We denote
the multivariate t-Student distribution as $\mathcal{T}$ parameterized
by mean, shape, and degree of freedom ($\mu,\Sigma,\nu$). We place
a spherical Gaussian prior on $\mu$ with zero mean and a Gamma prior
on $\nu$ with density concentrated in the range between 1 and 10.
We place an uninformative Wishart prior on $\Sigma$ as it is commonly
used on these tasks (\citet{leonard1992bayesian,alvarez2014bayesian}.
We perform mean-field variational inference, i.e. we consider an approximating
family which factorizes as:
\begin{equation}
q(\nu,\Sigma;\mu_{v},s,W,p,\alpha,\beta)=\mathcal{N}(\mu;\mu_{v},sI)\times\mathcal{W}(\Sigma^{-1};W,p)\times\Gamma(\nu;\alpha,\beta)
\end{equation}
where $\mathcal{W}$ denotes the Wishart distribution. This approximate
family is appropriate since it would be conditionally conjugate under
a Gaussian conditional model of the $\boldsymbol{x}_{i}$. More generally,
the Wishart distribution seems the natural choice of approximating
family for symmetric positive matrices because it has an explicit
density and we can sample from it. For $\mu_{v},s,W$ and $\beta$,
the reparameterization trick can be applied to obtain the gradient.
For $\alpha$ and $p$, we compare the impact of gradients estimated
by VIND and BBVI. While it would be possible to apply the GREP gradient
to $\alpha$, there is no simple approximate standardization available
for $\Sigma^{-1}$. We thus were unable to use the GREP gradient to
optimize $p$. For VIND, we use $\epsilon_{p}=2d$ and $\epsilon_{\alpha}=1$
for the variational parameters $p$ and $\alpha$, respectively.

\begin{figure}[h]
\centering{}\subfloat[ELBO on training period ($t=300$).]{\includegraphics[width=0.45\textwidth]{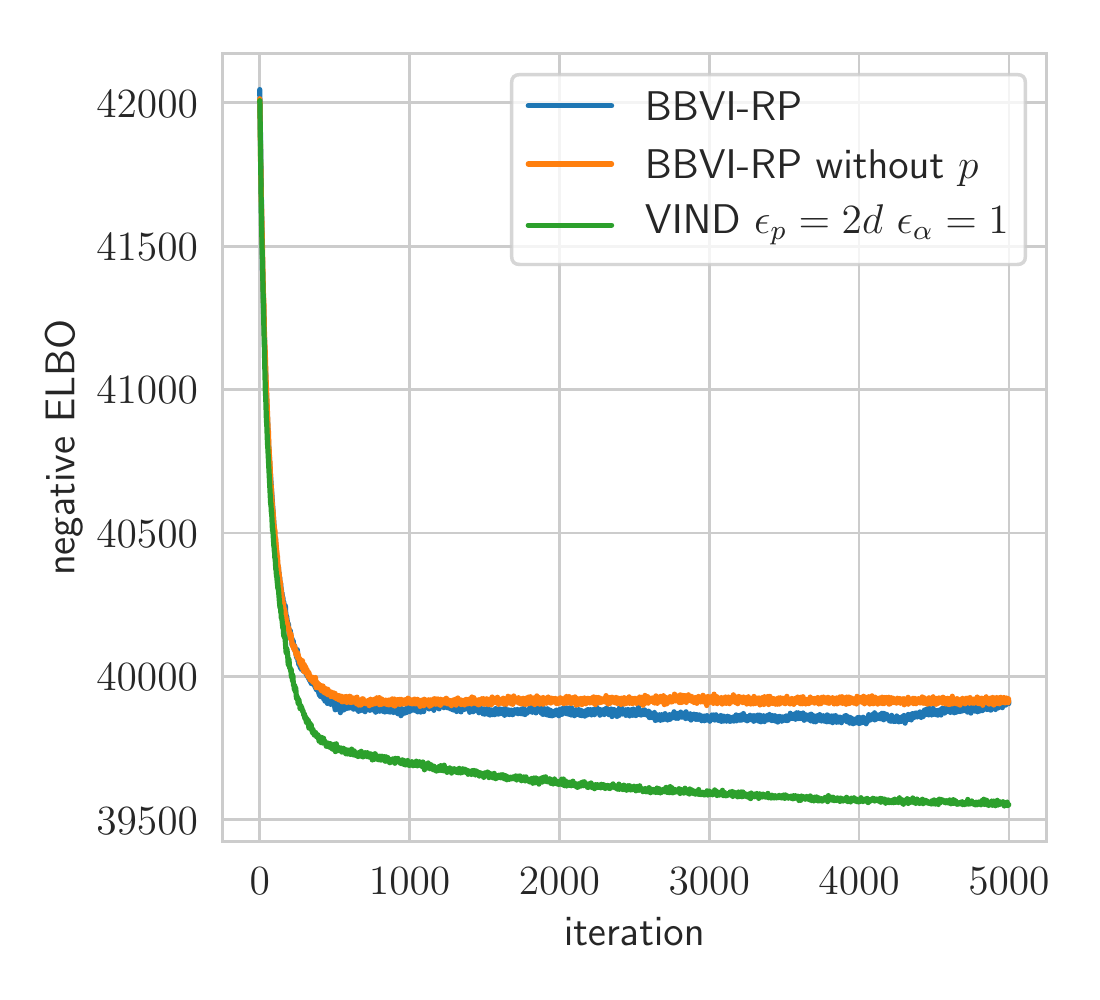}

}\hfill{}\subfloat[Log loss on held out period ($t=100$).]{\includegraphics[width=0.45\textwidth]{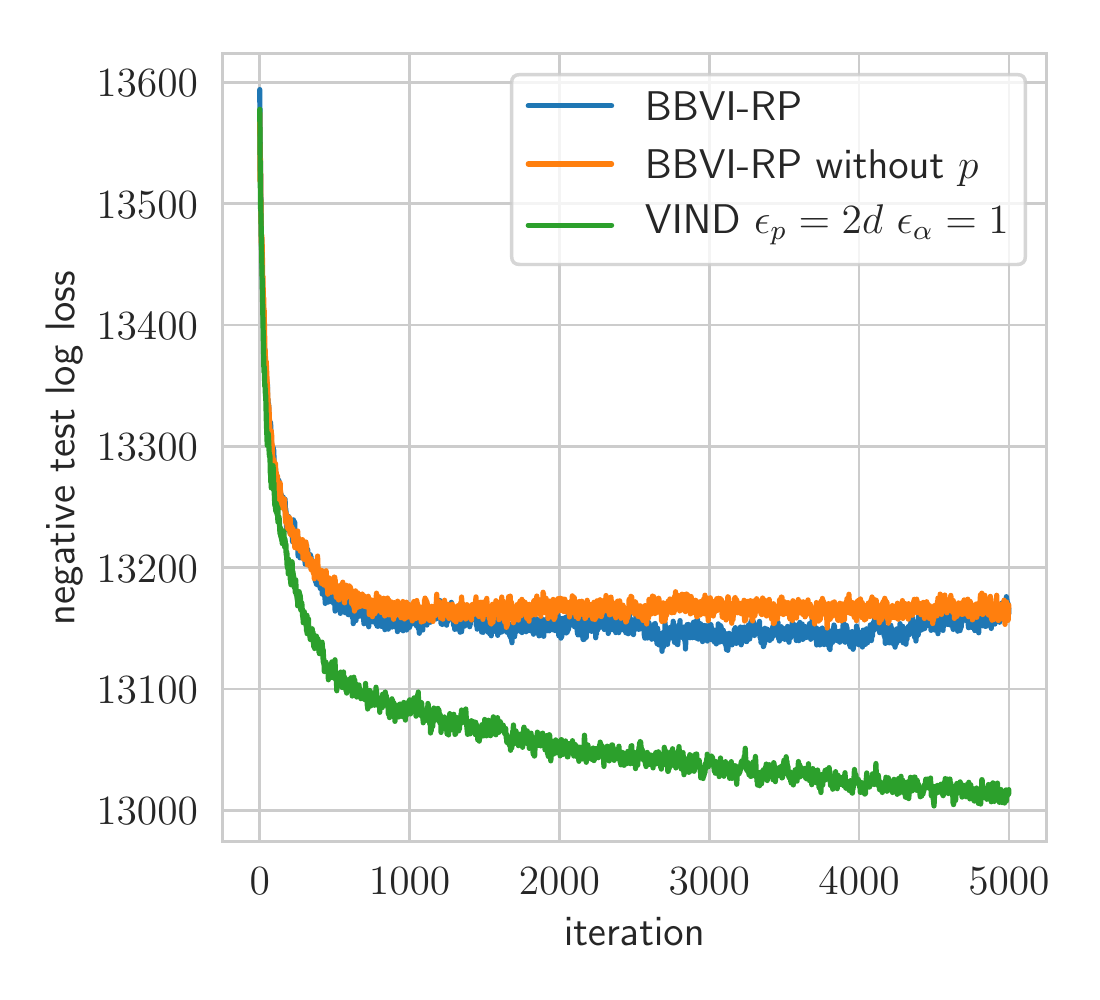}

}\caption{\textbf{Convergence of the negative ELBO on the financial location-scale
problem.}\protect \\
We plot the evolution of the negative-ELBO on the financial data model
for three Variational Inference algorithms. VIND (Green) and BBVI
(Blue) gradients are used on parameters for which reparameterization
cannot be achieved. BBVI without $p$ (Orange) refers to a BBVI optimization
of $\alpha$ with fixed value of the parameter $p.$ Until 100 iterations,
all algorithms have a similar ELBO due to this initial phase being
dominated by the convergence of the reparameterized parameters ($\mu_{v},s,\beta,W)$.
Due to a lower variance gradient in $(p,\alpha)$, VIND can properly
fit the approximation and attain a higher ELBO. In contrast, BBVI
hardly improves over freezing $p$ at its initial value. This is also
reflected on the test period. \label{fig:Convergence-behavior-on multivariate location-scale data}}
\end{figure}

Fig. \ref{fig:Convergence-behavior-on multivariate location-scale data}
shows the convergence behavior of both methods. $W$ is initialized
to the conjugate parameter according to a Wishart-Normal model while
all other parameters are initialized with the prior parameters. This
ensures that the initial approximate posterior of $\Sigma$ is concentrated
around the empirical covariance of the data. Both methods are applied
with three samples per iteration, which is the minimum for VIND.

Due to the reduced variance of the VIND gradient, we can attain a
higher ELBO on this task. Note that the BBVI's gradient of $p$ is
so noisy that it hardly improves the ELBO compared to keeping $p$
fixed (Fig.\ref{fig:Convergence-behavior-on multivariate location-scale data}).
The BBVI method thus completely fails in this example since it is
unable to tune the variability of the Wishart approximation of the
posterior distribution of $\Sigma^{-1}$. The computation time of
VIND and BBVI is the same. We further measure the variance of the
VIND and BBVI gradient (Appendix Section \ref{subsec:Wishart-Student-Financial}),
which shows that VIND exhibits gradients that have orders of magnitude
less variance just like in the synthetic experiment (Section \ref{sec: synthetic data}).

\section{Discussion}

From our experiments, it appears that VIND provides an efficient way
to find the best variational approximation in families for which the
reparameterization trick is inapplicable but for which there exists
a coupling. While it does require identifying a tight coupling of
$\boldsymbol{\theta}_{\dots\lambda_{i}\pm\epsilon}\dots$, we have
provided such couplings for a wide variety of approximation families
(Appendix Section \ref{sec:Couplings}). The VIND gradient thus provides
a major improvement over the BBVI baseline that remains competitive
with the GREP gradient (\citet{ruiz2016generalized}) while having
a much simpler derivation.

A key limit of VIND is that it scales poorly when it is used to compute
the gradient over many parameters, a key limit of numerical approximation
of gradients. This restricts its usage to approximating families for
which most parameters are amenable to reparameterization and for which
VIND is only used for a handful of key parameters. This is a key feature
of the Gamma and Wishart distributions for which only one scalar parameter
cannot be reparameterized. Indeed, in order to evaluate a gradient
against $p$ parameters using numerical derivatives, we need to evaluate
the function at $2p+1$ positions. Constructing a coupling on this
$2p+1$ space will require many more random samples than the usage
of the BBVI gradient. In approximating families with a large number
of parameters that are not amenable to reparameterization, the GREP
gradient from \citet{ruiz2016generalized} should thus be preferred.

However, we believe that this should be sufficient in the context
of simple statistical models (i.e. excluding Variational Auto-Encoders
and other models for which the posterior has a very complicated structure).
Indeed, if we have sufficiently many datapoints then, under mild assumptions
on the data and model, we should expect from a heuristic interpretation
of the Bernstein-von Mises theorem (\citet{van2000asymptotic,kleijn2012bernstein})
that the posterior distribution has a Gaussian limit. The family of
Gaussians thus has theoretical backing that other approximations lack
while also being amenable to reparameterization in all of its parameters.
Since it wins on both the theoretical and computational fronts, it
thus seems sensible to use the family of Gaussians as a default approximation
from which we should only deviate in key parameters, such as when
a conjugate prior would exist in a simpler model of the data as was
the case in our financial data application.

We have presented in this article how to apply VIND when the approximating
family is either Gamma, Beta, Dirichlet, Wishart, univariate and multivariate
Student or Poisson (Appendix Section \ref{sec:Couplings}). VIND is
thus applicable to a wide variety of critical approximation families.
Extending the algorithm further will be the subject of additional
work, but this seems to be a hard task since it requires the identification
of simple yet tight couplings of the various members of the family.

The key idea of VIND, using a coupling to minimize the variance of
the stochastic gradient, might also provide an interesting alternative
to existing algorithms for the maximization of the multi-sample ELBO
(\citet{cremer2017reinterpreting,domke2018importance}). Another possible
extension for VIND is the integration of additional variance reduction
techniques such as control variates (\citet{geffner2018using}) or
Rao-Blackwellization (\citet{ranganath2014black}). We will investigate
this in further research.

\bibliographystyle{plainnat}
\bibliography{ref.bib}

\newpage{}

\appendix

\section{Details of experiments\label{subsec:Further-Experiments}}

In this appendix, additional experimental results and details are
listed. Apart from further details on the experiments in the main
text, we present another benchmark using a standard linear regression
model, a real-world extension of the synthetic Gamma-Normal model
(Section \ref{sec: synthetic data}). 

\subsection{Synthetic Gamma-Normal Model\label{subsec:Synthetic-Gamma-Normal}}

\begin{figure}[h]
\subfloat[Smoothed bias of VIND-$\epsilon$, GREP, and BBVI gradient estimators.
Due to the noise in the BBVI gradient, we do not observe the expected
bias of a finite difference gradient compared to it.]{\begin{centering}
\includegraphics[width=0.45\textwidth]{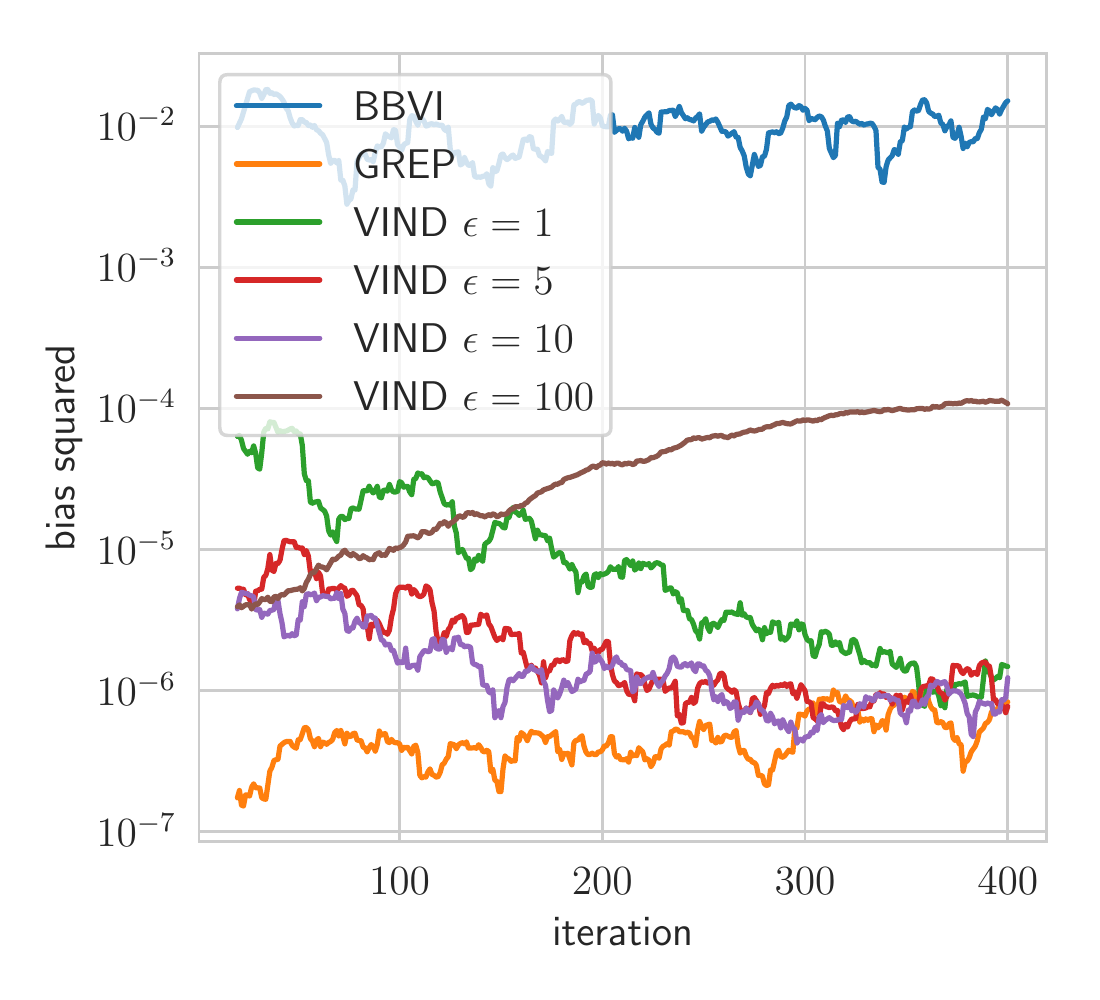}
\par\end{centering}
}\hfill{}\subfloat[Variance of VIND-$\epsilon$, GREP, and BBVI gradient estimators.
VIND provides less noisy gradients than BBVI for all displayed values
of $\epsilon$ and is comparable to GREP in later iterations, except
for $\epsilon=100$. The variance is almost equal to the MSE displayed
in main text Fig.\ref{fig: 1 MSE} but varies slightly for $\epsilon=100$
as the bias starts to have an effect.]{\centering{}\includegraphics[width=0.45\textwidth]{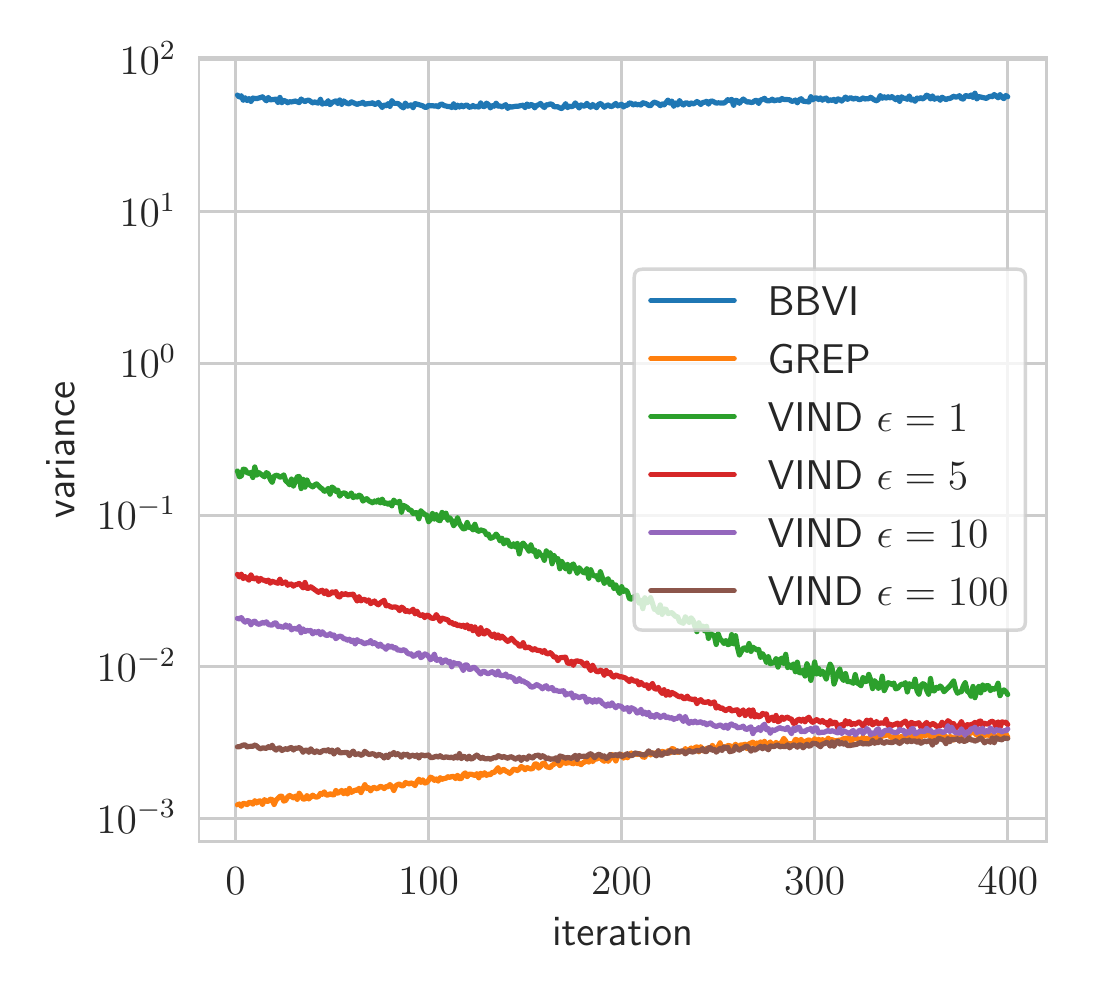}}

\caption{Bias (a) and variance (b) of various VIND, GREP and BBVI gradient
approximations along a descent following the actual gradient of the
ELBO. The estimates of bias has been smoothed with a window of 10
measurements for visual clarity. The variance is orders of magnitude
higher than the bias and hence dominates the MSE. The finite difference
gradients do not show signs of high bias even for the quite large
value of $\epsilon=10$. \label{fig:Bias-and-variance Toy model}}
\end{figure}

In Section \ref{sec: synthetic data}, we showed that the MSE of the
estimated gradient can be much lower using VIND compared to BBVI.
Due to the use of a conjugate model, we can compute both the variance
and bias of the estimators. We claimed that despite the finite difference
approach, VIND shows almost no bias and the variance of the estimated
gradient dominates the MSE. Figure \ref{fig:Bias-and-variance Toy model}
displays the variance and bias along the ascent path of the true gradient
of the ELBO in the Gamma-Normal experiment. Due to the high variance
of the BBVI gradient, the estimated bias is actually higher than that
of some VIND estimators. The variance is at least four orders of magnitude
larger than the bias, which supports our hypothesis that the eventual
bias of VIND is negligible and the variance reduction has a high impact.

\subsection{Wishart Student Financial Application\label{subsec:Wishart-Student-Financial}}

\begin{figure}[h]
\begin{centering}
\includegraphics[width=0.45\textwidth]{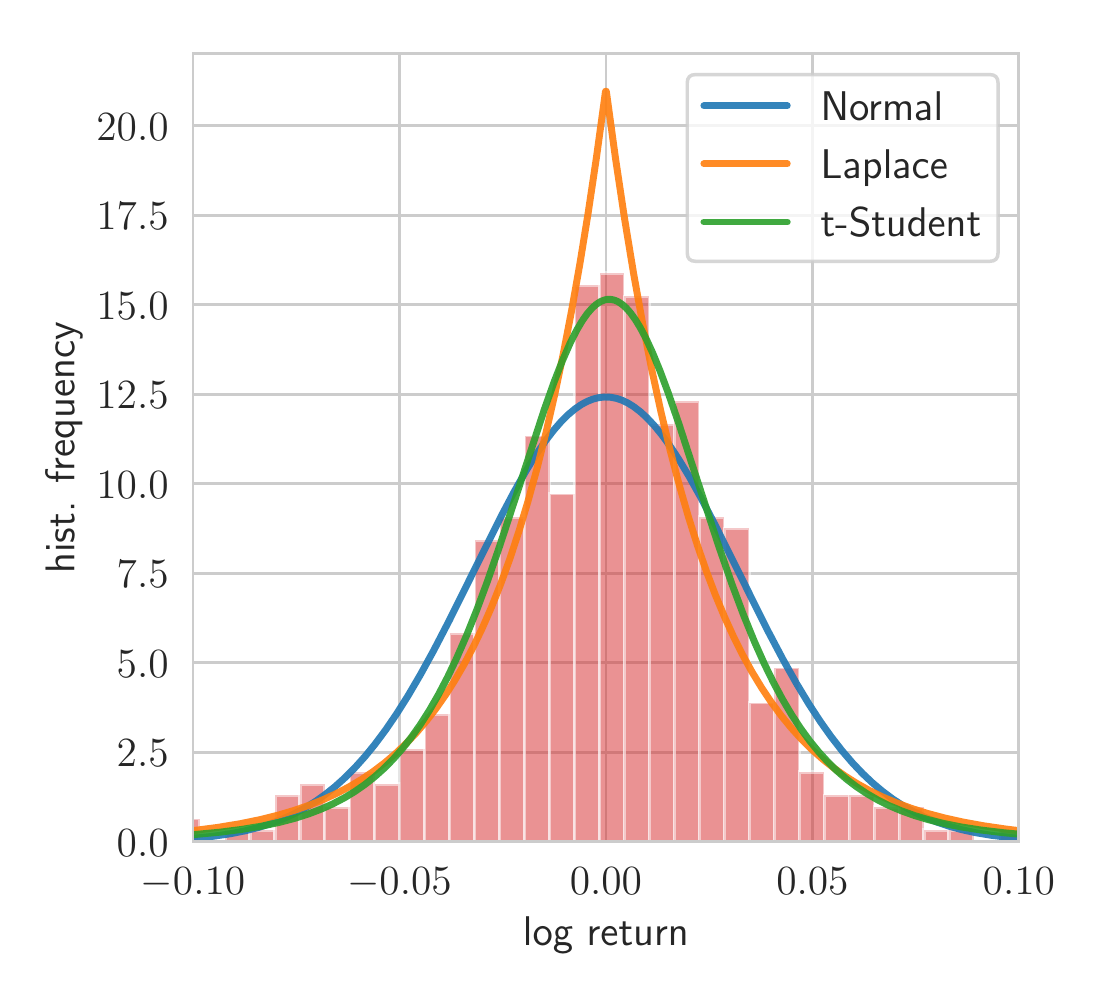}
\par\end{centering}
\caption{Maximum likelihood estimates of the Normal, Laplace, and t-Student
distributions on the log returns of CVX. The tails of the Normal do
not fit the data while the Laplace distribution is too spiky. The
additional degree-of-freedom parameter of the t-Student distribution
allows matching the tails well.\label{fig: normal model doesnt fit}}
\end{figure}

We provide additional details of our financial application. The symbols
of the stocks used are the following: \emph{AAPL, AXP, BA, CAT, CSCO,
CVX, DIS, DWDP, GE, GS, HD, IBM, INTC, JNJ, JPM, KO, MCD, MMM, MRK,
MSFT, NKE, PFE, TRV, UNH, UTX, V, VZ, WMT, }and \emph{XOM.} These
stocks are 29 of 30 companies in the Dow Jones as of May 2018. The
conjugate Gaussian model cannot provide a sufficiently close fit to
the data, as illustrated in Figure \ref{fig: normal model doesnt fit}.
This figure shows the histogram for 510 weeks of data and the maximum
likelihood fits using the Normal, Laplace, and t-Student distribution.
The Laplace and Normal distributions fit poorly in the tails whereas
the t-Student can fit the shape of the histogram due to the additional
degree of freedom.

As pointed out in Section \ref{subsec:Wishart-Student-Financial},
we start all variational parameters from the prior parameters except
for $W.$ For $W,$ we compute the empirical covariance of the data,
invert it and divide it by the initial parameter for $p.$ The expected
value of the Wishart distribution is thus the empirical precision
matrix. This reduces the number of iterations until convergence and
highlights the final convergence. The results are reproducible with
different optimization procedures (e.g. Adagrad) and various initialization
parameters. BBVI can only attain a comparable maximum ELBO if the
degree of freedom of the posterior Wishart distribution $p$ is initialized
to the optimum reached using VIND.

Similar to the synthetic experiment presented in Section \ref{fig:Bias-and-variance Toy model},
we estimate the variance of different gradient estimators for this
model. Having no access to the true gradient, we can only estimate
the variance. Figure \ref{fig:Estimated-variance-of VIND, VIND_uncoupled, BBVI}
shows the variance of VIND, VIND without coupling, and BBVI with Rao-Blackwellization
along the descent path of VIND. The gradient estimates using BBVI
have a much higher variance than those of VIND. VIND without coupling
has approximately one order of magnitude higher variance but is still
a better solution than BBVI on this problem. The high gradient variance
leads to the inability to properly converge in the variational parameter
$p$ which is apparent in Figure \ref{fig:Convergence-behavior-on multivariate location-scale data}
in Section \ref{subsec:Synthetic-Gamma-Normal}.

\begin{figure}
\begin{centering}
\includegraphics[width=0.45\textwidth]{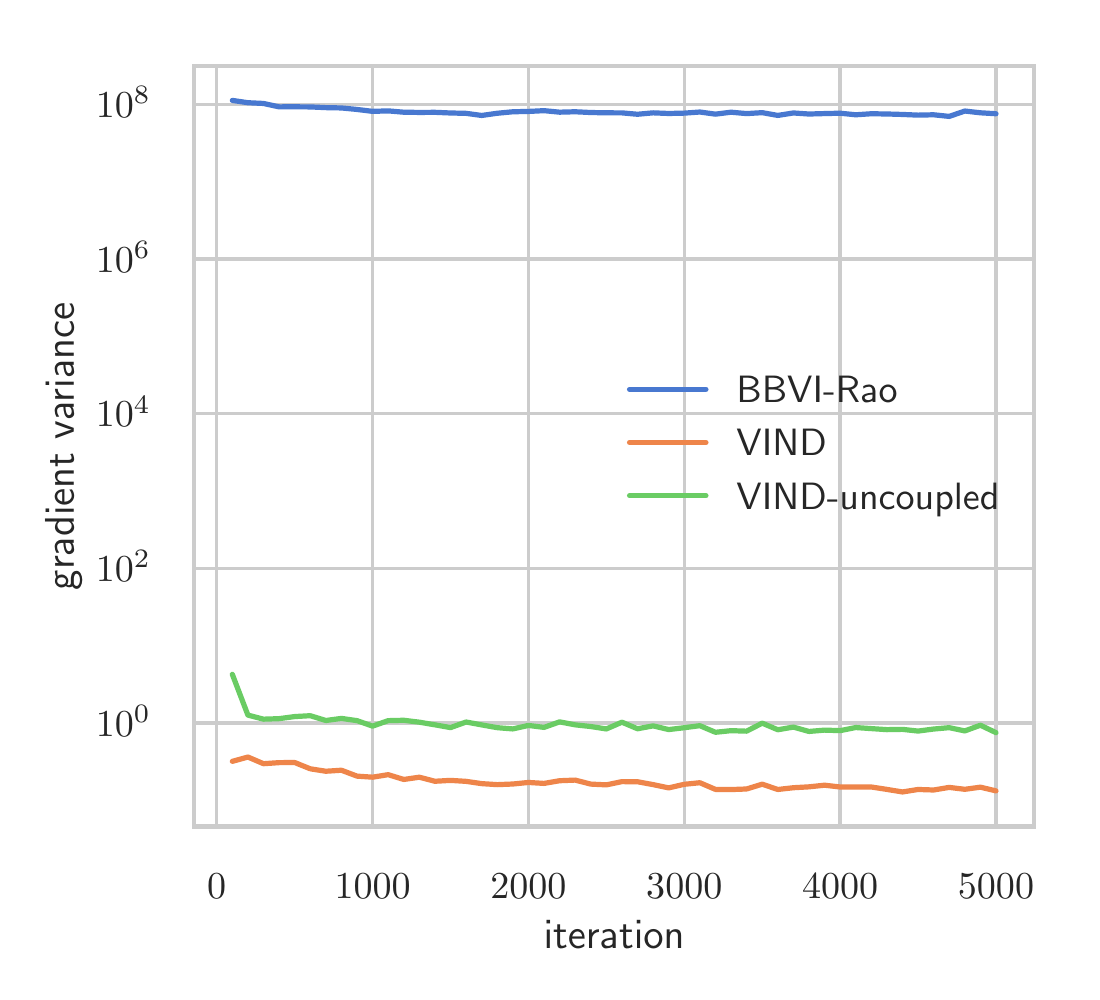}
\par\end{centering}
\caption{Estimated variance of VIND and BBVI gradients. The BBVI gradient uses
the variance-reduction technique of Rao-Blackwellization (\citet{ranganath2014black}).
The VIND algorithm is presented with and without coupling. The variance
is measured along ELBO ascent with 5000 samples using VIND. To estimate
the variance, we take 1000 samples every 100 iterations. VIND gradients
are much less noisy than the Rao-Blackwellized BBVI gradient.\label{fig:Estimated-variance-of VIND, VIND_uncoupled, BBVI}}
\end{figure}

\subsection{Linear Regression}

\begin{figure}
\subfloat[Negative ELBO during training.]{\centering{}\includegraphics[width=0.45\textwidth]{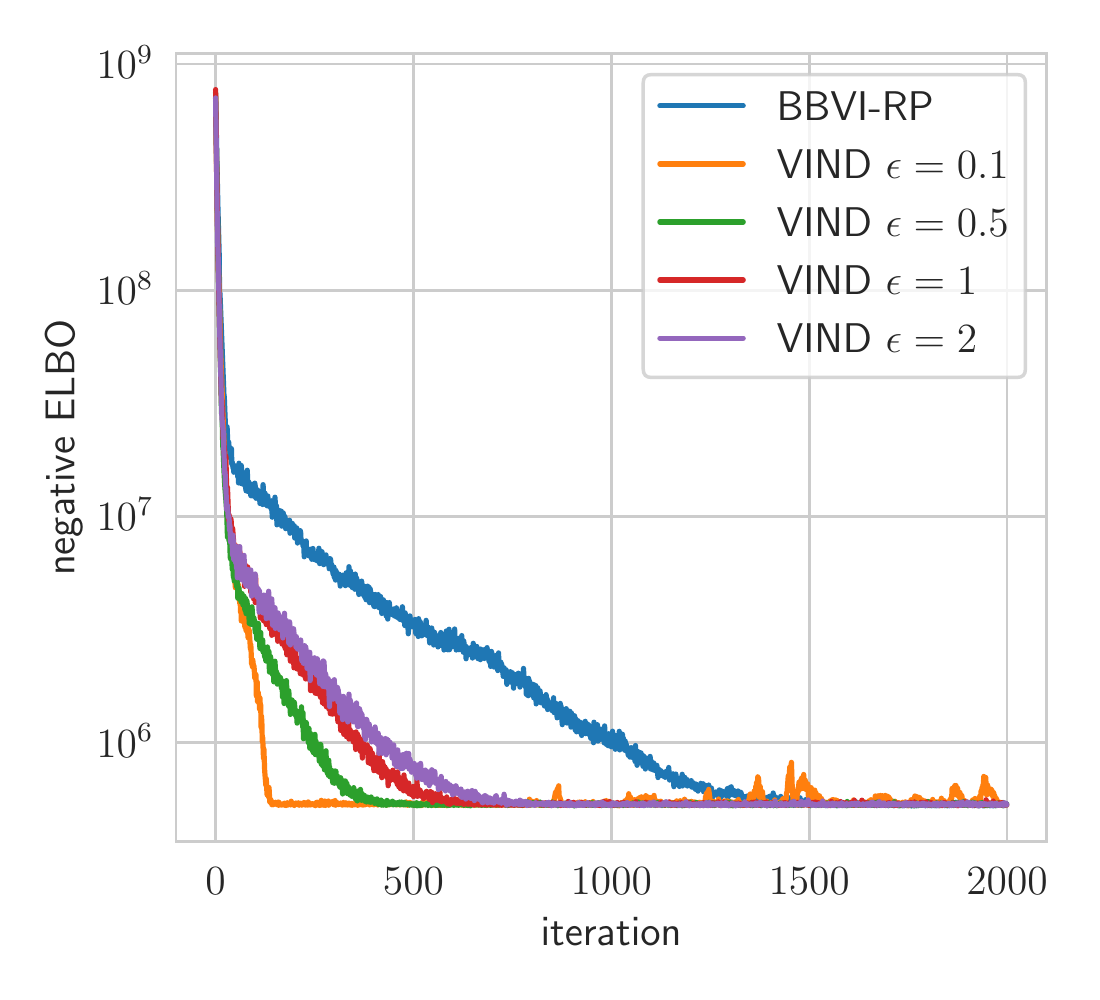}}\hfill{}\subfloat[Negative test log loss on a holdout set.]{\centering{}\includegraphics[width=0.45\textwidth]{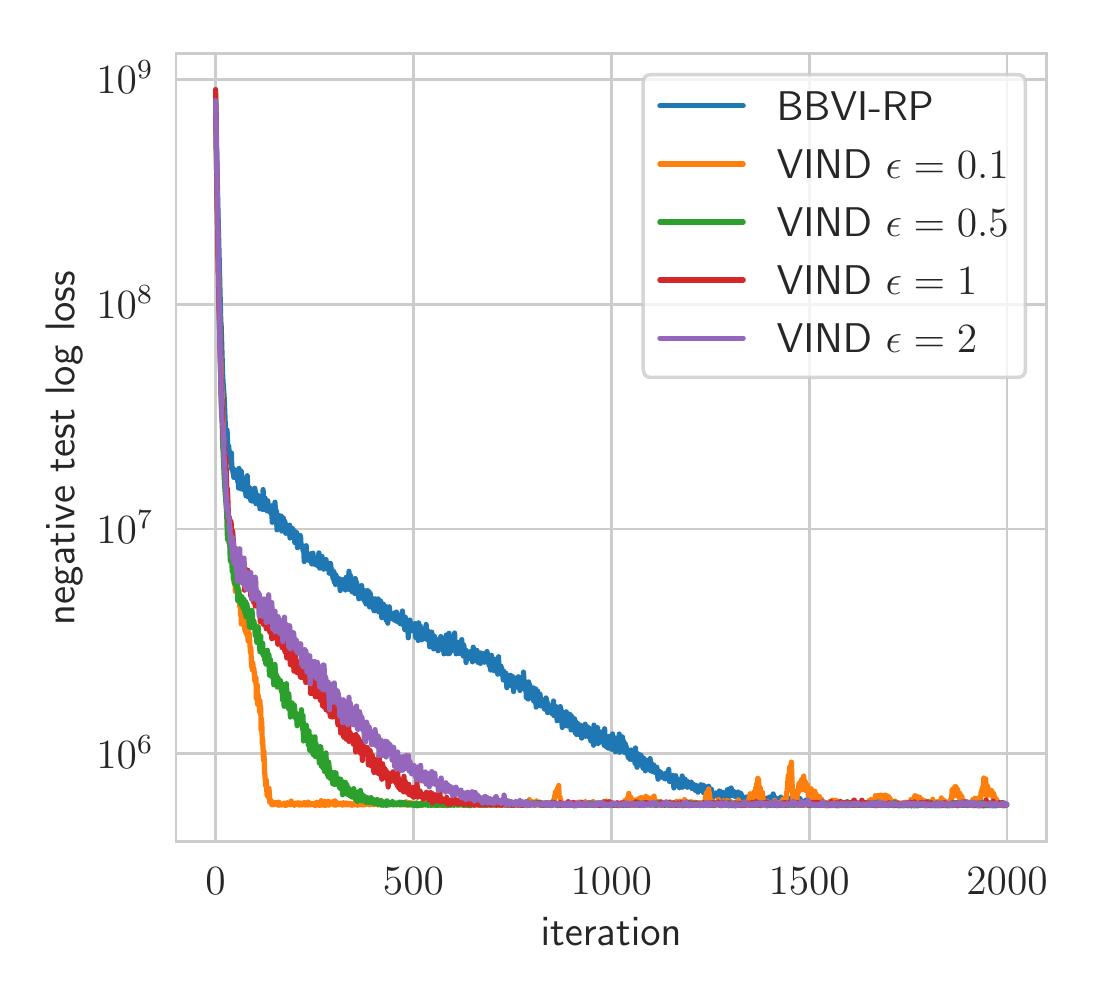}}
\centering{}\caption{Convergence behavior on the linear regression task with independent
prior and mean-field approximating family. Negative ELBO and test
log-loss compared for VIND with different values of $\epsilon$and
BBVI. VIND with $\epsilon=1$ converges after around 500 iterations
while BBVI requires 2500. The same minimum is attained by both methods.\label{fig: linear regression}}
\end{figure}

The synthetic example with a Gamma-Normal model has shown that VIND
can estimate gradients with much lower variance compared to BBVI.
A common application of the Gamma distribution arises in a linear
regression setting. We observe a set of feature vectors paired with
a real quantity to predict, i.e. we have the data set $\mathcal{X}=\{x_{i}\}_{i=1}^{n}$
and $\mathcal{Y}=\{y_{i}\}_{i=1}^{n}$ with $x_{i}\in\mathbb{R}^{d}$
and $y_{i}\in\mathbb{R}$. We use the Boston Housing benchmark data
set with $n=506$ and $d=13$. We center and standardize the features
and further decorrelate them using principal component analysis. We
have the following conditional model and prior, respectively:

\[
f(\mathcal{Y}|w,\tau,\mathcal{X})=\prod_{i=1}^{n}\mathcal{N}(y_{i}|x_{i}^{T}w,\tau^{-1})\qquad p(w,\tau)=\mathcal{N}(w;0,s_{0}I)\times\Gamma(\tau;\alpha_{0},\beta_{0}).
\]
We would have a conjugate model if the prior was hierarchical, i.e.
$p(w,\tau)=p(\tau)p(w|\tau)$. With the independent prior, we need
to resort to approximate methods such as variational inference. The
approximating family is of the same structure as the prior (mean-field)
and, with $\mu,s\in\mathbb{R}^{d}$, is given by

\[
q(w,\tau;\mu,s,\alpha,\beta)=\mathcal{N}(w;\mu,\text{diag}(s))\times\Gamma(\tau;\alpha,\beta).
\]
The prior parameters are $\alpha_{0}=\beta_{0}=5,s_{0}=1$ and we
have a cold start on all parameters with $\alpha=200,\beta=50,s=1$.
Figure \ref{fig: linear regression} depicts the convergence of the
negative ELBO for VIND with different values of epsilon and BBVI.
The procedure converges much faster using the gradient estimated with
VIND. Using $\epsilon=1$, VIND converges after around 500 iterations
while BBVI requires approximately 2500. In comparison to the Wishart-Student
model analyzed in Section \ref{subsec:Wishart-Student-Financial},
BBVI can attain the same loss as VIND but requires much more time
to do so.

\pagebreak{}

\section{Derivation of the VIND update\label{sec:Derivation-of-the VIND}}

Let us now derive in detail the precise form of the VIND gradient
and why it differs from the immediate numerical approximation of the
gradient of the ELBO.

Throughout this section, we assume that we are dealing with a one-dimensional
parameter approximation family: $q\left(\theta;\lambda\right)$. The
argument is straightforward to extend to the higher-dimensional case.

The numerical approximation of the derivative of the ELBO is:
\begin{align}
\widetilde{\nabla}_{\lambda}ELBO & =\frac{ELBO(\lambda+\epsilon)-ELBO(\lambda-\epsilon)}{2\epsilon}\\
 & =\frac{1}{2\epsilon}\mathbb{E}\left[\log\frac{p\left(\theta_{\lambda+\epsilon}\right)}{q\left(\theta_{\lambda+\epsilon};\lambda+\epsilon\right)}-\log\frac{p\left(\theta_{\lambda-\epsilon}\right)}{q\left(\theta_{\lambda-\epsilon};\lambda-\epsilon\right)}\right]
\end{align}
In the limit $\epsilon\rightarrow0$, a Taylor expansion of the ELBO
centered at $\lambda$ yields that this approximation is exact up
to order $\mathcal{O}\left(\epsilon^{2}\right)$ if the ELBO is differentiable
three times:
\begin{align}
ELBO\left(\lambda+\epsilon\right) & =ELBO\left(\lambda\right)+\epsilon\nabla_{\lambda}ELBO\left(\lambda\right)+\frac{\epsilon^{2}}{2}H_{\lambda}ELBO\left(\lambda\right)+\mathcal{O}\left(\epsilon^{3}\right)\\
ELBO\left(\lambda-\epsilon\right) & =ELBO\left(\lambda\right)-\epsilon\nabla_{\lambda}ELBO\left(\lambda\right)+\frac{\epsilon^{2}}{2}H_{\lambda}ELBO\left(\lambda\right)+\mathcal{O}\left(\epsilon^{3}\right)\\
\frac{ELBO(\lambda+\epsilon)-ELBO(\lambda-\epsilon)}{2\epsilon} & =0+\frac{2\epsilon}{2\epsilon}\nabla_{\lambda}ELBO\left(\lambda\right)+0+\frac{1}{2\epsilon}\mathcal{O}\left(\epsilon^{3}\right)\\
\widetilde{\nabla}_{\lambda}ELBO & =\nabla_{\lambda}ELBO\left(\lambda\right)+\mathcal{O}\left(\epsilon^{2}\right)
\end{align}

However, the naive numerical approximation of the derivative is slightly
suboptimal. This can be shown by a Taylor expansion of $\log q\left(\theta;\lambda\pm\epsilon\right)$
around $\lambda$. For all $\theta$, we have:
\begin{align}
\log q\left(\theta;\lambda+\epsilon\right)-\log q\left(\theta;\lambda-\epsilon\right) & =0+2\epsilon\nabla_{\lambda}\log q\left(\theta;\lambda\right)+0+\mathcal{O}\left(\epsilon^{3}\right)
\end{align}
We thus have:
\begin{align}
 & \frac{1}{2\epsilon}\mathbb{E}\left[\log\frac{p\left(\theta_{\lambda+\epsilon}\right)}{q\left(\theta_{\lambda+\epsilon};\lambda+\epsilon\right)}-\log\frac{p\left(\theta_{\lambda-\epsilon}\right)}{q\left(\theta_{\lambda-\epsilon};\lambda-\epsilon\right)}\right]\nonumber \\
 & =\frac{1}{2\epsilon}\mathbb{E}\left[\log\frac{p\left(\theta_{\lambda+\epsilon}\right)}{q\left(\theta_{\lambda+\epsilon};\lambda\right)}-\log\frac{p\left(\theta_{\lambda-\epsilon}\right)}{q\left(\theta_{\lambda-\epsilon};\lambda\right)}\right]+\mathbb{E}\left[\nabla_{\lambda}\log q\left(\theta;\lambda\right)+\mathcal{O}\left(\epsilon^{2}\right)\right]\\
 & =\nabla_{\lambda,VIND}ELBO+\mathbb{E}\left[\nabla_{\lambda}\log q\left(\theta;\lambda\right)\right]+\mathcal{O}\left(\epsilon^{2}\right)
\end{align}

The extra term is actually equal to $\mathcal{O}\left(\epsilon^{2}\right)$
since the expected value of the gradient is $0$:
\begin{align}
\mathbb{E}\left[\nabla_{\lambda}\log q\left(\theta;\lambda\right)\right] & =\int\nabla_{\lambda}\left[\log q\left(\theta;\lambda\right)\right]q\left(\theta;\lambda\right)d\theta\\
 & =\int\frac{\nabla_{\lambda}\left[q\left(\theta;\lambda\right)\right]}{q\left(\theta;\lambda\right)}q\left(\theta;\lambda\right)d\theta\\
 & =\int\nabla_{\lambda}\left[q\left(\theta;\lambda\right)\right]d\theta\\
 & =\nabla_{\lambda}\left[\int q\left(\theta;\lambda\right)d\theta\right]\\
 & =\nabla_{\lambda}\left[1\right]\\
 & =0
\end{align}

We have thus established that the difference between the true numerical
approximation of the derivative of the ELBO (eq.\eqref{eq: Finite differences approximation of the gradient})
and the VIND approximation (eq.\eqref{eq: VIND gradient}) corresponds
to the numerical approximation of a term with derivative $0$. It
is thus more efficient to use the VIND approximation of the gradient
which deliberately sets this term exactly to $0$. This is compounded
by the result in the next Section which establishes that the VIND
update coincides with the reparameterization gradient in a transformation
family. This would not be true for the true numerical approximation
of the derivative of the ELBO in which there is an extra term present
which is exactly $\mathbb{E}\left[\nabla_{\lambda}\log q\left(\theta;\lambda\right)\right]$.

\section{Linking VIND to the reparameterization gradient\label{sec:Linking-VIND-to}}

If $q\left(\boldsymbol{\theta};\lambda\right)$ is a transformation
family, then the VIND gradient is almost equal to the reparameterization
gradient, as we now show.

First, let us build the straightforward coupling of $\boldsymbol{\theta}_{\lambda+\epsilon}$
and $\boldsymbol{\theta}_{\lambda-\epsilon}$. We do so by using the
transformation family property:
\begin{equation}
\boldsymbol{\theta}_{\lambda\pm\epsilon}=F\left(\boldsymbol{Z},\lambda\pm\epsilon\right)
\end{equation}
where the distribution of $\boldsymbol{Z}$ is fixed.

The VIND approximation of the gradient of the ELBO is then:
\begin{align}
\nabla_{\lambda,VIND}ELBO & =\frac{1}{2\epsilon}\mathbb{E}\left(\log\frac{p\left(\boldsymbol{\theta}_{\lambda+\epsilon}\right)}{q\left(\boldsymbol{\theta}_{\lambda+\epsilon};\lambda\right)}-\log\frac{p\left(\boldsymbol{\theta}_{\lambda-\epsilon}\right)}{q\left(\boldsymbol{\theta}_{\lambda-\epsilon};\lambda\right)}\right)\\
 & =\frac{1}{2\epsilon}\mathbb{E}\left(\log\frac{p\left(F\left(\boldsymbol{Z},\lambda+\epsilon\right)\right)}{q\left(F\left(\boldsymbol{Z},\lambda+\epsilon\right);\lambda\right)}-\log\frac{p\left(F\left(\boldsymbol{Z},\lambda-\epsilon\right)\right)}{q\left(F\left(\boldsymbol{Z},\lambda-\epsilon\right);\lambda\right)}\right)
\end{align}
which simplifies in the limit $\epsilon\rightarrow0$. Indeed:
\begin{align}
\log p\left(F\left(\boldsymbol{Z},\lambda+\epsilon\right)\right) & \approx\log p\left(F\left(\boldsymbol{Z},\lambda\right)\right)+\epsilon J_{\lambda}F\left(\boldsymbol{Z},\lambda\right)\nabla_{\boldsymbol{\theta}}\log p\left(F\left(\boldsymbol{Z},\lambda\right)\right)\\
\log p\left(F\left(\boldsymbol{Z},\lambda-\epsilon\right)\right) & \approx\log p\left(F\left(\boldsymbol{Z},\lambda\right)\right)-\epsilon J_{\lambda}F\left(\boldsymbol{Z},\lambda\right)\nabla_{\boldsymbol{\theta}}\log p\left(F\left(\boldsymbol{Z},\lambda\right)\right)\\
\log q\left(F\left(\boldsymbol{Z},\lambda+\epsilon\right);\lambda\right) & \approx\log q\left(F\left(\boldsymbol{Z},\lambda\right);\lambda\right)+\epsilon J_{\lambda}F\left(\boldsymbol{Z},\lambda\right)\nabla_{\boldsymbol{\theta}}\log q\left(F\left(\boldsymbol{Z},\lambda\right);\lambda\right)\\
\log q\left(F\left(\boldsymbol{Z},\lambda-\epsilon\right);\lambda\right) & \approx\log q\left(F\left(\boldsymbol{Z},\lambda\right);\lambda\right)-\epsilon J_{\lambda}F\left(\boldsymbol{Z},\lambda\right)\nabla_{\boldsymbol{\theta}}\log q\left(F\left(\boldsymbol{Z},\lambda\right);\lambda\right)
\end{align}
Thus yielding:
\begin{align}
\log\frac{p\left(F\left(\boldsymbol{Z},\lambda+\epsilon\right)\right)}{q\left(F\left(\boldsymbol{Z},\lambda+\epsilon\right);\lambda\right)}-\log\frac{p\left(F\left(\boldsymbol{Z},\lambda-\epsilon\right)\right)}{q\left(F\left(\boldsymbol{Z},\lambda-\epsilon\right);\lambda\right)} & \approx2\epsilon\left(J_{\lambda}F\left(\boldsymbol{Z},\lambda\right)\nabla_{\boldsymbol{\theta}}\log\frac{p\left(F\left(\boldsymbol{Z},\lambda\right)\right)}{q\left(F\left(\boldsymbol{Z},\lambda\right);\lambda\right)}\right)
\end{align}

We thus finally have:
\begin{equation}
\lim_{\epsilon\rightarrow0}\nabla_{\lambda,VIND}ELBO=\mathbb{E}\left(J_{\lambda}F\left(\boldsymbol{Z};\lambda\right)\nabla_{\boldsymbol{\theta}}\log\frac{p\left(F\left(\boldsymbol{Z};\lambda\right)\right)}{q\left(F\left(\boldsymbol{Z};\lambda\right);\lambda\right)}\right)
\end{equation}
which is exactly the expression of the reparameterization gradient
of the ELBO.

\newpage{}

\section{Couplings\label{sec:Couplings}}

In this section, we detail how to derive the VIND gradient to find
an approximation of the target distribution $f\left(\boldsymbol{\theta}\right)$
in the Gamma, Beta, Dirichlet, Wishart, univariate and multivariate
Student or Poisson families. For each of those families, we detail
a coupling a construction of $\boldsymbol{\theta}_{\lambda}$ and
$\boldsymbol{\theta}_{\boldsymbol{\lambda}_{\pm\epsilon}}^{\left(j\right)}$
for all parameters which are not amenable to a reparameterization
and we give a reparameterization transform for all other parameters.

Please notice that, throughout this section, we do not take advantage
of the fact that the entropy of most of those distributions is explicit
(i.e. we use eq.\eqref{eq: ELBO definition} to define the ELBO).
All the following formulas thus have a variant where only the terms
in $p\left(\boldsymbol{\theta}\right)$ remain and the terms of depending
on $q\left(\boldsymbol{\theta}\right)$ are replaced by an explicit
gradient $\nabla_{\boldsymbol{\lambda}}H_{q}\left(\boldsymbol{\lambda}\right)$,
based on eq.\eqref{eq: ELBO with entropy}. This alternative formula
could yield improvements on the variance of the estimator depending
on the circumstance through the use of control variates (\citet{geffner2018using}).

\subsection{Gamma distribution}

Consider the family of Gamma distributions. The conditional density
is:
\begin{equation}
q\left(\theta;\alpha,\beta\right)\propto\theta^{\alpha-1}\exp\left(-\beta\theta\right)
\end{equation}
The $\beta$ parameter is an inverse-scale parameter and thus has
a standard reparameterization gradient. The $\alpha$ parameter needs
to be handled instead with a VIND gradient through the following coupling:\begin{subequations}
\begin{align}
\gamma_{\alpha-\epsilon} & \sim\Gamma\left(\alpha-\epsilon,1\right)\\
\gamma_{\epsilon,1} & \sim\Gamma\left(\epsilon,1\right)\\
\gamma_{\epsilon,2} & \sim\Gamma\left(\epsilon,1\right)\\
\theta_{\alpha-\epsilon,\beta} & =\frac{1}{\beta}\gamma_{\alpha-\epsilon}\\
\theta_{\alpha+\epsilon,\beta} & =\frac{1}{\beta}\left(\gamma_{\alpha-\epsilon}+\gamma_{\epsilon,1}+\gamma_{\epsilon,2}\right)\\
\theta_{\alpha,\beta} & =\frac{1}{\beta}\left(\gamma_{\alpha-\epsilon}+\gamma_{\epsilon,1}\right)
\end{align}
\end{subequations} This coupling leverages the key fact that a sum
of two Gamma random variables with the same scale parameter is also
a Gamma random variable. The alpha parameter of the sum is equal to
the sum of the alpha parameters of the summands.

Given $n$ independent samples from this coupling $\boldsymbol{\theta}^{\left(1\right)}\dots\boldsymbol{\theta}^{\left(n\right)}$,
the VIND gradient is: \begin{subequations} 
\begin{align}
\nabla_{\beta}ELBO & \approx\frac{1}{n}\left\{ \sum_{j=1}^{n}\left(-\frac{\gamma_{\alpha-\epsilon}+\gamma_{\epsilon,1}}{\beta^{2}}\right)\frac{\partial}{\partial\theta}\left[\log\frac{p\left(\theta_{\alpha,\beta}^{\left(j\right)}\right)}{q\left(\theta_{\alpha,\beta}^{\left(j\right)};\alpha,\beta\right)}\right]\right\} \\
\nabla_{\alpha,VIND}ELBO & \approx\frac{1}{2\epsilon}\frac{1}{n}\sum_{j=1}^{n}\left(\log\frac{p\left(\theta_{\alpha+\epsilon,\beta}^{\left(j\right)}\right)}{q\left(\theta_{\alpha+\epsilon,\beta}^{\left(j\right)};\alpha,\beta\right)}-\log\frac{p\left(\theta_{\alpha-\epsilon,\beta}^{\left(j\right)}\right)}{q\left(\theta_{\alpha-\epsilon,\beta}^{\left(j\right)};\alpha,\beta\right)}\right)
\end{align}
 \end{subequations} 

\subsection{Beta distribution}

Consider the family of Beta distributions. The conditional density
is:
\begin{equation}
q\left(\theta;\alpha,\beta\right)\propto\theta^{\alpha-1}\left(1-\theta\right)^{\beta-1}
\end{equation}
Both parameters need to be handled through a coupling construction.
We leverage here the key fact that a Beta random variable can be constructed
as the ratio of two gammas: \begin{subequations} 
\begin{align}
\gamma_{\alpha} & \sim\Gamma\left(\alpha,1\right)\\
\gamma_{\beta} & \sim\Gamma\left(\beta,1\right)\\
\beta & =\frac{\gamma_{\alpha}}{\gamma_{\alpha}+\gamma_{\beta}}\\
 & \sim\mathcal{B}\left(\alpha,\beta\right)
\end{align}
 \end{subequations} From this, we can construct a coupling of $\theta_{\alpha\pm\epsilon,\beta}$
and $\theta_{\alpha,\beta\pm\epsilon}$ in the following way. First,
we construct $\theta_{\alpha,\beta}$: \begin{subequations} 
\begin{align}
\gamma_{\alpha-\epsilon} & \sim\Gamma\left(\alpha-\epsilon,1\right)\\
\gamma_{\epsilon,1} & \sim\Gamma\left(\epsilon,1\right)\\
\gamma_{\alpha} & =\gamma_{\alpha-\epsilon}+\gamma_{\epsilon,1}\\
\gamma_{\beta-\epsilon} & \sim\Gamma\left(\beta-\epsilon,1\right)\\
\gamma_{\epsilon,2} & \sim\Gamma\left(\epsilon,1\right)\\
\gamma_{\beta} & =\gamma_{\beta-\epsilon}+\gamma_{\epsilon,2}\\
\theta_{\alpha,\beta} & =\frac{\gamma_{\alpha}}{\gamma_{\alpha}+\gamma_{\beta}}
\end{align}
and then we construct the perturbed versions:
\begin{align}
\gamma_{\epsilon,3} & \sim\Gamma\left(\epsilon,1\right)\\
\gamma_{\alpha+\epsilon,\beta} & =\gamma_{\alpha}+\gamma_{\epsilon,3}\\
\gamma_{\epsilon,4} & \sim\Gamma\left(\epsilon,1\right)\\
\gamma_{\beta+\epsilon} & =\gamma_{\beta}+\gamma_{\epsilon,4}\\
\theta_{\alpha-\epsilon,\beta} & =\frac{\gamma_{\alpha-\epsilon}}{\gamma_{\alpha-\epsilon}+\gamma_{\beta}}\\
\theta_{\alpha+\epsilon,\beta} & =\frac{\gamma_{\alpha+\epsilon}}{\gamma_{\alpha+\epsilon}+\gamma_{\beta}}\\
\theta_{\alpha,\beta-\epsilon} & =\frac{\gamma_{\alpha}}{\gamma_{\alpha}+\gamma_{\beta-\epsilon}}\\
\theta_{\alpha,\beta+\epsilon} & =\frac{\gamma_{\alpha}}{\gamma_{\alpha}+\gamma_{\beta+\epsilon}}
\end{align}
 \end{subequations} 

This construction yields the following VIND gradient: \begin{subequations} 
\begin{align}
\nabla_{\alpha,VIND}ELBO & \approx\frac{1}{2\epsilon}\frac{1}{n}\sum_{j=1}^{n}\left(\log\frac{p\left(\theta_{\alpha+\epsilon,\beta}^{\left(j\right)}\right)}{q\left(\theta_{\alpha+\epsilon,\beta}^{\left(j\right)};\alpha,\beta\right)}-\log\frac{p\left(\theta_{\alpha-\epsilon,\beta}^{\left(j\right)}\right)}{q\left(\theta_{\alpha-\epsilon,\beta}^{\left(j\right)};\alpha,\beta\right)}\right)\\
\nabla_{\beta,VIND}ELBO & \approx\frac{1}{2\epsilon}\frac{1}{n}\sum_{j=1}^{n}\left(\log\frac{p\left(\theta_{\alpha,\beta+\epsilon}^{\left(j\right)}\right)}{q\left(\theta_{\alpha,\beta+\epsilon}^{\left(j\right)};\alpha,\beta\right)}-\log\frac{p\left(\theta_{\alpha,\beta-\epsilon}^{\left(j\right)}\right)}{q\left(\theta_{\alpha,\beta-\epsilon}^{\left(j\right)};\alpha,\beta\right)}\right)
\end{align}
 \end{subequations} 

\subsection{Dirichlet distribution}

The Dirichlet distribution is a distribution over a $p$-dimensional
random variable $\boldsymbol{\theta}$ with density:
\begin{equation}
q\left(\boldsymbol{\theta};\alpha_{1}\dots\alpha_{p}\right)\propto\left[\prod_{i=1}^{p}\left(\theta_{i}\right)^{\alpha_{i}-1}\right]\boldsymbol{1}\left(\sum_{i=1}^{p}\theta_{i}=1\text{ and }\forall_{i}\ \theta_{i}\geq0\right)
\end{equation}
It is supported on the $p$-dimensional simplex defined as the ensemble
of points such that $\sum_{i=1}^{p}\theta_{i}=1$ and $\theta_{i}\geq0$
for all coordinates $i$.

Exactly like the Beta distribution, the Dirichlet distribution can
be constructed as a ratio of Gamma random variables. More precisely,
given $p$ Gamma random variables:
\begin{equation}
\gamma_{i}\sim\Gamma\left(\alpha_{i},1\right)
\end{equation}
then the random variable with coordinates:
\begin{equation}
\theta_{i}=\frac{\gamma_{i}}{\sum_{k=1}^{p}\gamma_{k}}
\end{equation}
follows a Dirichlet distribution with parameters $\alpha_{1}\dots\alpha_{k}$.

Thus, we can construct a coupling by first generating $3p$ Gamma
random variables: \begin{subequations} 
\begin{align}
\gamma_{i,\alpha_{i}-\epsilon} & \sim\Gamma\left(\alpha_{i}-\epsilon,1\right)\\
\gamma_{i,1} & \sim\Gamma\left(\epsilon,1\right)\\
\gamma_{i,2} & \sim\Gamma\left(\epsilon,1\right)\\
\gamma_{i,\alpha} & =\gamma_{i,\alpha_{i}-\epsilon}+\gamma_{i,1}\\
\gamma_{i,\alpha+\epsilon} & =\gamma_{i,\alpha_{i}}+\gamma_{i,2}
\end{align}
 \end{subequations} The variable $\boldsymbol{\theta}_{\alpha_{1}\dots\alpha_{n}}$
and its distorted variants are then straightforward to compute: \begin{subequations} 
\begin{align}
\left(\boldsymbol{\theta}_{\alpha_{1}\dots\alpha_{n}}\right)_{i} & =\frac{\gamma_{i,\alpha}}{\sum_{k=1}^{p}\gamma_{k,\alpha}}\\
\left(\boldsymbol{\theta}_{\alpha_{1}\dots\alpha_{j\pm\epsilon}\dots\alpha_{n}}\right)_{i} & =\frac{\boldsymbol{1}\left(i\neq j\right)\gamma_{i,\alpha}+\boldsymbol{1}\left(i=j\right)\gamma_{j,\alpha\pm\epsilon}}{\gamma_{j,\alpha\pm\epsilon}+\sum_{k\neq j}\gamma_{k,\alpha}}
\end{align}
 \end{subequations} 

The VIND gradient is then straightforward to derive and matches that
of the Beta distribution.

\subsection{Wishart distribution}

The Wishart distribution is a distribution over the space of positive
matrices of shape $\left(p,p\right)$. It has two parameters: the
degree of freedom $d\geq p$ and the scale matrix $\boldsymbol{V}$
which is a strictly positive matrix of shape $\left(p,p\right)$.
Let $\boldsymbol{S}$ denote a Wishart random variable. Its density
is:
\begin{equation}
q\left(\boldsymbol{S};d,\boldsymbol{V}\right)\propto\left|\boldsymbol{S}\right|^{\left(d-p-1\right)/2}\exp\left(-\frac{1}{2}\text{tr}\left(\boldsymbol{V}^{-1}\boldsymbol{S}\right)\right)
\end{equation}
where $\left|\boldsymbol{S}\right|$ is the determinant of the matrix
$\boldsymbol{S}$ and $\text{tr}\left(\cdot\right)$ is the trace
operator.

The Wishart distribution is the matrix equivalent of the Gamma distribution.
The $\boldsymbol{V}$ parameter can be handled through a reparameterization
while the degree of freedom parameter $d$ requires a coupling construction.
The coupling construction is made possible by the fact that the sum
of two Wishart random variables with identical scale $\boldsymbol{V}$
but different degrees of freedom $d_{1}$ and $d_{2}$ is another
Wishart variable with scale $\boldsymbol{V}$ and degree of freedom
$d_{1}+d_{2}$.

Let $\boldsymbol{V}=\boldsymbol{C}^{2}$ where $\boldsymbol{C}$ is
a symmetric matrix. We propose the following coupling: \begin{subequations} 
\begin{align}
\boldsymbol{W}_{1} & \sim\mathcal{W}\left(d-\epsilon,\boldsymbol{I}_{p}\right)\\
\boldsymbol{W}_{2} & \sim\mathcal{W}\left(\epsilon,\boldsymbol{I}_{p}\right)\\
\boldsymbol{W}_{3} & \sim\mathcal{W}\left(\epsilon,\boldsymbol{I}_{p}\right)\\
\boldsymbol{S}_{d-\epsilon,\boldsymbol{V}} & =\boldsymbol{C}\boldsymbol{W}_{1}\boldsymbol{C}\\
\boldsymbol{S}_{d,\boldsymbol{V}} & =\boldsymbol{C}\left(\boldsymbol{W}_{1}+\boldsymbol{W}_{2}\right)\boldsymbol{C}\\
\boldsymbol{S}_{d+\epsilon,\boldsymbol{V}} & =\boldsymbol{C}\left(\boldsymbol{W}_{1}+\boldsymbol{W}_{2}+\boldsymbol{W}_{3}\right)\boldsymbol{C}
\end{align}
 \end{subequations} 

Given $n$ samples from this coupling $\boldsymbol{W}^{\left(1\right)}\dots\boldsymbol{W}^{\left(n\right)}$,
we have the following VIND gradient: \begin{subequations} 
\begin{align}
\nabla_{\boldsymbol{V}}ELBO & \approx\frac{1}{n}\sum_{j=1}^{n}\nabla_{\boldsymbol{C}}\left[\boldsymbol{C}\left(\boldsymbol{W}_{1}+\boldsymbol{W}_{2}\right)\boldsymbol{C}\right]\nabla_{\boldsymbol{S}}\left[\log\frac{p\left(\boldsymbol{S}_{d,\boldsymbol{V}}^{\left(j\right)}\right)}{q\left(\boldsymbol{S}_{d,\boldsymbol{V}}^{\left(j\right)};d,\boldsymbol{V}\right)}\right]\label{eq: reparam grad V in Wishart}\\
\nabla_{d,VIND}ELBO & \approx\frac{1}{2\epsilon}\frac{1}{n}\sum_{j=1}^{n}\left[\log\frac{p\left(\boldsymbol{S}_{d+\epsilon,\boldsymbol{V}}^{\left(j\right)}\right)}{q\left(\boldsymbol{S}_{d+\epsilon,\boldsymbol{V}}^{\left(j\right)};d,\boldsymbol{V}\right)}-\log\frac{p\left(\boldsymbol{S}_{d-\epsilon,\boldsymbol{V}}^{\left(j\right)}\right)}{q\left(\boldsymbol{S}_{d-\epsilon,\boldsymbol{V}}^{\left(j\right)};d,\boldsymbol{V}\right)}\right]
\end{align}
 \end{subequations} Please notice that the gradients against $\boldsymbol{S}$
and against $\boldsymbol{C}$ in eq.\eqref{eq: reparam grad V in Wishart}
need to take into account that both matrices are symmetric.

\subsection{Univariate and multivariate Student}

The multivariate Student distribution with center $\boldsymbol{\mu}$,
scale $\boldsymbol{S}$ (a symmetric matrix), and degree of freedom
$d$ over $\mathbb{R}^{p}$ has the following density:
\begin{equation}
q\left(\boldsymbol{\theta};d,\boldsymbol{\mu},\boldsymbol{S}\right)\propto\left[1+\frac{1}{d}\left(\boldsymbol{\theta}-\boldsymbol{\mu}\right)^{T}\left(\boldsymbol{S}\right)^{-2}\left(\boldsymbol{\theta}-\boldsymbol{\mu}\right)\right]^{-\left(d+p\right)/2}
\end{equation}
A key property of this distribution is that it corresponds to the
marginal distribution of a scaled and translated ratio of a Standard
normal and a chi-square variable: \begin{subequations} 
\begin{align}
\boldsymbol{Z} & \sim\mathcal{N}\left(0,\boldsymbol{I}_{p}\right)\\
c & \sim\chi_{d}^{2}\\
\boldsymbol{\theta} & =\boldsymbol{\mu}+\boldsymbol{S}\frac{\boldsymbol{Z}}{\sqrt{c/d}}
\end{align}
 \end{subequations} The marginals of this distribution are, once
we center and scale them appropriately, the ordinary Student distribution.

The $\boldsymbol{\mu}$ and $\boldsymbol{S}$ parameters correspond
to a center and a scale and are thus amenable to reparameterization.
The $d$ parameter requires a coupling construction: \begin{subequations} 
\begin{align}
c_{d-\epsilon} & \sim\chi_{d-\epsilon}^{2}\\
c_{\epsilon,1} & \sim\chi_{\epsilon}^{2}\\
c_{\epsilon,2} & \sim\chi_{\epsilon}^{2}\\
\boldsymbol{Z} & \sim\mathcal{N}\left(0,\boldsymbol{I}_{p}\right)\\
\boldsymbol{\theta}_{d,\boldsymbol{\mu},\boldsymbol{S}} & =\boldsymbol{\mu}+\boldsymbol{S}\frac{\boldsymbol{Z}}{\sqrt{\left(c_{d-\epsilon}+c_{\epsilon,1}\right)/d}}\\
\boldsymbol{\theta}_{d+\epsilon,\boldsymbol{\mu},\boldsymbol{S}} & =\boldsymbol{\mu}+\boldsymbol{S}\frac{\boldsymbol{Z}}{\sqrt{\left(c_{d-\epsilon}+c_{\epsilon,1}+c_{\epsilon,2}\right)/d}}\\
\boldsymbol{\theta}_{d-\epsilon,\boldsymbol{\mu},\boldsymbol{S}} & =\boldsymbol{\mu}+\boldsymbol{S}\frac{\boldsymbol{Z}}{\sqrt{\left(c_{d-\epsilon}\right)/d}}
\end{align}
 \end{subequations} which yields the VIND gradient: \begin{subequations} 
\begin{align}
\nabla_{\boldsymbol{\mu}}ELBO & \approx\frac{1}{n}\sum_{j=1}^{n}\nabla_{\boldsymbol{\theta}}\frac{\log p\left(\boldsymbol{\theta}_{d,\boldsymbol{\mu},\boldsymbol{S}}^{\left(j\right)}\right)}{\log q\left(\boldsymbol{\theta}_{d,\boldsymbol{\mu},\boldsymbol{S}}^{\left(j\right)};d,\boldsymbol{\mu},\boldsymbol{S}\right)}\\
\nabla_{\boldsymbol{S}}ELBO & \approx\frac{1}{n}\sum_{j=1}^{n}\Bigg[\frac{1}{2}\frac{\left(\boldsymbol{Z}^{\left(j\right)}\right)}{\sqrt{\left(c_{d-\epsilon}^{\left(j\right)}+c_{\epsilon,1}^{\left(j\right)}\right)/d}}\nabla_{\boldsymbol{\theta}}^{T}\frac{\log p\left(\boldsymbol{\theta}_{d,\boldsymbol{\mu},\boldsymbol{S}}^{\left(j\right)}\right)}{\log q\left(\boldsymbol{\theta}_{d,\boldsymbol{\mu},\boldsymbol{S}}^{\left(j\right)};d,\boldsymbol{\mu},\boldsymbol{S}\right)}\nonumber \\
 & \ \ \ \ +\frac{1}{2}\nabla_{\boldsymbol{\theta}}\frac{\log p\left(\boldsymbol{\theta}_{d,\boldsymbol{\mu},\boldsymbol{S}}^{\left(j\right)}\right)}{\log q\left(\boldsymbol{\theta}_{d,\boldsymbol{\mu},\boldsymbol{S}}^{\left(j\right)};d,\boldsymbol{\mu},\boldsymbol{S}\right)}\frac{\left(\boldsymbol{Z}^{\left(j\right)}\right)^{T}}{\sqrt{\left(c_{d-\epsilon}^{\left(j\right)}+c_{\epsilon,1}^{\left(j\right)}\right)/d}}\Bigg]\\
\nabla_{d,VIND}ELBO & \approx\frac{1}{2\epsilon}\frac{1}{n}\sum_{j=1}^{n}\left[\frac{\log p\left(\boldsymbol{\theta}_{d+\epsilon,\boldsymbol{\mu},\boldsymbol{S}}^{\left(j\right)}\right)}{\log q\left(\boldsymbol{\theta}_{d+\epsilon,\boldsymbol{\mu},\boldsymbol{S}}^{\left(j\right)};d,\boldsymbol{\mu},\boldsymbol{S}\right)}-\frac{\log p\left(\boldsymbol{\theta}_{d-\epsilon,\boldsymbol{\mu},\boldsymbol{S}}^{\left(j\right)}\right)}{\log q\left(\boldsymbol{\theta}_{d-\epsilon,\boldsymbol{\mu},\boldsymbol{S}}^{\left(j\right)};d,\boldsymbol{\mu},\boldsymbol{S}\right)}\right]
\end{align}
 \end{subequations} 

\subsection{Poisson distribution}

The Poisson distribution is a discrete distribution over $\mathbb{N}$
with mass function:
\begin{equation}
f\left(k;\lambda\right)=\frac{\lambda^{k}}{k!}\exp\left(-\lambda\right)
\end{equation}
It has the key property that the sum of two Poisson random variables
with rates $\lambda_{1}$ and $\lambda_{2}$ is another Poisson random
variable with rate $\lambda_{1}+\lambda_{2}$.

This yields the coupling: \begin{subequations} 
\begin{align}
k_{\lambda-\epsilon} & \sim\mathcal{P}\left(\lambda-\epsilon\right)\\
k_{2\epsilon} & \sim\mathcal{P}\left(2\epsilon\right)\\
k_{\lambda+\epsilon} & =k_{\lambda-\epsilon}+k_{2\epsilon}
\end{align}
 \end{subequations} 

This yields the naive VIND gradient:
\begin{equation}
\nabla_{\lambda,VIND}ELBO\approx\frac{1}{n}\sum_{j=1}^{n}\frac{1}{2\epsilon}\left[\log\frac{p\left(k_{\lambda+\epsilon}\right)}{q\left(k_{\lambda+\epsilon};\lambda\right)}-\log\frac{p\left(k_{\lambda-\epsilon}\right)}{q\left(k_{\lambda-\epsilon};\lambda\right)}\right]
\end{equation}
However, this naive gradient can be improved. Indeed, it is inefficient
due to the high probability of the event $k_{2\epsilon}=0$ which
yields a null gradient. It is better to first modify the VIND gradient
by explicitly conditioning on the events $k_{2\epsilon}=0$ and $k_{2\epsilon}\neq0$:
\begin{align}
\nabla_{\lambda,VIND}ELBO & =\frac{1}{2\epsilon}\mathbb{E}\left[\log\frac{p\left(k_{\lambda+\epsilon}\right)}{q\left(k_{\lambda+\epsilon};\lambda\right)}-\log\frac{p\left(k_{\lambda-\epsilon}\right)}{q\left(k_{\lambda-\epsilon};\lambda\right)}\right]\\
 & =\frac{1}{2\epsilon}\left\{ 0+\mathbb{P}\left(k_{2\epsilon}\neq0\right)\mathbb{E}\left[\log\frac{p\left(k_{\lambda+\epsilon}\right)}{q\left(k_{\lambda+\epsilon};\lambda\right)}-\log\frac{p\left(k_{\lambda-\epsilon}\right)}{q\left(k_{\lambda-\epsilon};\lambda\right)}|k_{2\epsilon}\neq0\right]\right\} \\
 & =\frac{1-\exp\left(-2\epsilon\right)}{2\epsilon}\mathbb{E}\left[\log\frac{p\left(k_{\lambda+\epsilon}\right)}{q\left(k_{\lambda+\epsilon};\lambda\right)}-\log\frac{p\left(k_{\lambda-\epsilon}\right)}{q\left(k_{\lambda-\epsilon};\lambda\right)}|k_{2\epsilon}\neq0\right]
\end{align}
which is straightforward to evaluate with a sampling approximation.
We simply construct $n$ samples $\tilde{k}^{\left(1\right)}\dots\tilde{k}^{\left(n\right)}$
using the conditional distribution of $k_{2\epsilon}$ when $k_{2\epsilon}\neq0$:
\begin{equation}
\nabla_{\lambda,VIND}ELBO\approx\frac{1-\exp\left(-2\epsilon\right)}{2\epsilon}\frac{1}{n}\sum_{j=1}^{n}\left[\log\frac{p\left(\tilde{k}_{\lambda+\epsilon}\right)}{q\left(\tilde{k}_{\lambda+\epsilon};\lambda\right)}-\log\frac{p\left(\tilde{k}_{\lambda-\epsilon}\right)}{q\left(\tilde{k}_{\lambda-\epsilon};\lambda\right)}\right]
\end{equation}
This improved VIND gradient is more efficient since it avoids sampling
from the $k_{2\epsilon}=0$ event which provides no information concerning
the sign of the gradient.
\end{document}